\documentclass[apj,numberedappendix]{emulateapj}

\usepackage{color}
\usepackage{enumitem}
\usepackage{hyperref}
\usepackage{verbatim}
\usepackage{amsmath,amstext,mathrsfs}
\usepackage[all]{hypcap} 
\usepackage{afterpage}
\usepackage{graphicx}

\begin{document}
\title{Tracing magnetic fields by the synergies of synchrotron emission gradients }

\author{Jian-Fu Zhang\altaffilmark{1,3}, Qian Liu\altaffilmark{2}, Alex Lazarian\altaffilmark{3}} 
\email{jfzhang@xtu.edu.cn}
\altaffiltext{1}{Department of Physics, Xiangtan University, Xiangtan, Hunan 411105, China}
\altaffiltext{2}{School of Languages and Media, Anhui University of Finance and Economics, Bengbu, Anhui  233030, China}
\altaffiltext{3}{Astronomy Department, University of Wisconsin, Madison, WI 53711, USA}
\begin{abstract}
This paper studies how to employ synchrotron emission gradient techniques to reveal the properties of the magnetic field within the interstellar media. Based on data cubes of three-dimensional numerical simulations of magnetohydrodynamic turbulence, we explore spatial gradients of synchrotron emission diagnostics to trace the direction of the magnetic field. According to our simulations, multifarious diagnostics for synchrotron emission can effectively determine the potential direction of projected magnetic fields. Applying the synergies of synchrotron diagnostic gradients to the archive data from the Canadian Galactic Plane Survey, we find that multifarious diagnostic techniques make consistent predictions for the Galactic magnetic field directions. With the high-resolution data presently available from 
Low Frequency Array for radio astronomy and those in the future from the Square Kilometer Array, the synergies of synchrotron emission gradients are supposed to perform better in tracing the actual direction of interstellar magnetic fields, especially in the low-frequency Faraday rotation regime where traditional synchrotron polarization measure fails.
\end{abstract}


\section{Introduction}\label{intro}
Magnetohydrodynamic (MHD) turbulence is extremely ubiquitous throughout the interstellar medium (ISM, \citealt{Armstrong95} ). In light of its important role
in key astrophysical processes, such as star formation, propagation and acceleration of cosmic rays, heat transport, as well as magnetic reconnection (see \citealt{Lazarian99}, henceforth LV99; \citealt{Narayan01,Elmegreen04,MacLow04,Mckee07}), the further comprehension of MHD turbulence properties, a necessary prerequisite for the study of these astrophysical problems, will provide timely help to reveal the inherent nature of astrophysical processes.

Synchrotron polarization signal, from the non-thermal relativistic electrons spiraling in magnetized turbulent media, can uncover information of turbulence, which suggests that relating observational statistics of synchrotron fluctuation with underlying statistics of magnetic fields will be a promising approach to investigating MHD turbulence. Traditionally, the polarization angle ($\theta={1\over 2} {\rm arctan}(U/Q)$), in which Stokes parameters $Q$ and $U$ representing two perpendicular directions respectively can provide an intuitive description of the polarization signal, can be employed to characterize the direction of linearly polarized synchrotron emission. However, the lack of correctness can be found in this polarization vector method tracing magnetic field directions in the Faraday depolarization regime and limitations appear in the case of interferometric data such as missing single-dish data in the $Q$-$U$ plane. 

For years polarized synchrotron fluctuations together with the Faraday rotation measure, known as the Faraday rotation synthesis or Faraday tomography (\citealt{Burn66,Brentjens05}), have been popular in astrophysical community. This technique sheds new light on the magnetic structure of the Milky Way and neighboring galaxies (e.g., \citealt{Frick11,Beck13,Jelic15,VanEck17}). Valuable information as it provides on the plane-of-sky geometry of the magnetic field for external galaxies, little information has been obtained in magnetic field directions along the line of sight (LOS) because sign changes can happen in the LOS component of magnetic fields, resulting in the disorder of Faraday depth. Apparently, the ambiguous estimation of the LOS magnetic field lies in the framework of the traditional Faraday tomography (\citealt{Ferriere16}).

Synchrotron polarization gradient technique, an innovative approach proposed by \cite{Gaensler11}, draws attention in investigation of the interstellar turbulence. This approach introduces a rotationally and translationally invariant quantity, that is the magnitude of \textit{gradients of the complex polarization vector} of $\textit{\textbf{P}}=Q+iU$, to help diagnose the properties of the ISM.\footnote{In case of confusion, the bold notation $\textit{\textbf{P}}$ denotes the vector quantity of the linear polarization.} In this diagnostic, it is true that the images present the complex network of tangled filamentary structures and statistical analyses constrain the sonic Mach number of the turbulence within the warm ionized ISM (\citealt{Gaensler11}), but little can be done to probe neutral gas turbulence (\citealt{Herron17}).

On the basis of statistical descriptions of any random variable, a series of the formulation of synchrotron polarization intensity fluctuations was proposed by Lazarian \& Pogosyan (2016, hereafter LP16) to recover the magnetic turbulence, and tested by \cite{Zhang16,Zhang18} with synthetic simulations. Those numerical studies demonstrated that simulations are in agreement with the theoretical predictions made in LP16. Based on previous theoretical predictions and present understanding of MHD turbulence (e.g.,  \citealt{Gold95}, henceforth GS95), the new synchrotron gradient techniques, synchrotron intensity (\textit{scalar $I$}), polarization intensity (\textit{scalar $P$}) and its derivative to squared wavelength (\textit{scalar $dP\over d\lambda^2$}) included, were suggested by Lazarian \& Yuen (2018, henceforth LY18) to trace magnetic field directions and reconstruct a 3D structure of turbulent magnetic fields. Very recently, \cite{Zhang19} has advanced synchrotron polarization gradient techniques by highlighting the multi-frequency alignment measure (AM) and extending the previous work into super-Alfv{\'e}nic turbulence regime. One of the main findings is that in the low frequency and strong Faraday rotation range, synchrotron gradient techniques take a significant advantage over the traditional polarization method in tracing projected magnetic fields. 

With the purpose of excluding the effect of artifacts from missing single-dish data in the $Q$-$U$ plane, \cite{Herron18a} first suggested other advanced diagnostics, including the generalized polarization gradient, polarization directional curvature, polarization wavelength derivative and polarization wavelength curvature, and then 
tested both simulation and observational images of diffuse synchrotron polarization emission (\citealt{Herron18b}). The authors were not aware of the relation between the magnetic field direction and gradients of synchrotron intensities and polarization intensities discovered on the basis of the analysis of the fundamental properties of magnetic turbulence and turbulent reconnection in a series of papers that introduced a Synchrotron Gradient Technique (SGT) of tracing magnetic field using synchrotron observations (Lazarian et al. 2017, LY18). Therefore they did not explore whether the new measures they introduced can trace magnetic field in observations. In this paper we extend the SGT by exploring the relative ability of synchrotron polarization measures introduced in Herron et al. (2018b) to trace the magnetic field direction. For this purpose we use both data cubes obtained with 3D compressible MHD simulations as well as the Canadian Galactic Plane Survey (CGPS). 

The paper is organized as follows. In Section 2, we give brief descriptions regarding the theory of MHD turbulence, synchrotron emission diagnostics and gradient measure technique. Section 3 presents the results from simulation data cubes, followed by the application of the gradient measure to the CGPS data in Section 4.  
Discussion and summary are presented in Sections 5 and 6, respectively.

\section{Method Descriptions}
\subsection{Modern theory of MHD turbulence}\label{MuoMT}
The modern theory of MHD turbulence was proposed in the GS95 outstanding work, which presents a collection of anisotropic eddies aligning with the direction of magnetic fields surrounding them. Later analytical and numerical investigations extended theoretical descriptions with emergence of some new concepts: the definition of the local system of frame (\citealt{ChoV00}) and treatment of compressibility of MHD turbulence (\citealt{Cho02PRL}), to name just a few. The alignment of the eddies to the local direction of magnetic fields follows the theory of turbulent magnetic reconnection (LV99), which predicts that the reconnection in turbulent fluid occurs over just one eddy turnover period. Not only do magnetic fields promote the motion of eddies mixing up magnetic field lines, but also offer the eddies more energy with less resistance in the situation of random driving. As a result, most of the kinetic energy of Alfv{\'e}nic motion is concentrated in such eddies that are perpendicular to the local direction of magnetic fields. This alignment between anisotropic eddies and local magnetic fields is confirmed by numerical simulations (\citealt{ChoV00,Maron01}). 

Only when the energy of magnetic fields over the volume of eddies is larger than or comparable with the kinetic energy of eddies are the magnetic fields important for regulating the direction of the eddies. The relationship between the energy of magnetic fields and the eddies is determined by the Alfv{\'e}nic Mach number, i.e., $M_{\rm A}=V_{\rm L}/V_{\rm A}$, where $V_{\rm L}$ is the injection velocity of turbulence driving at the scale $L_{\rm inj}$ and $V_{\rm A}= B/\sqrt{4\pi \rho}$ is an Alfv{\'e}nic velocity in a plasma with the density $\rho$. It should be emphasized that the GS95 work with far-reaching implications focuses on a strong trans-Alfv{\'e}nic ($M_{\rm A}\sim 1$) incompressible turbulence mode. Under the condition of a critical balance, i.e., $v_{l}l_{\perp}^{-1}=V_{\rm A}l_{\|}^{-1}$, where $v_ l$ is the fluctuation velocity at the scale $l$, while $l_{\|}$ and $l_{\perp}$ refer to parallel and perpendicular scales of the eddies, respectively, the derived relationship of 
\begin{equation}
l_{\|}\propto l_{\perp}^{2/3} \label{eq:aniso}
\end{equation}
works for characterizing anisotropies of the turbulence. The applicability of the theory of realistic compressible turbulence was obtained in \cite{Lithwick01} and \cite{Cho02PRL}. The latter demonstrated that fluctuations of Alfv{\'e}n and slow modes for both the velocities and magnetic fields make a marginal difference while fluctuations of density are seriously modified by compressibility. 

What's more, the GS95 work was generalized to both $M_{\rm A}<1$ and $M_{\rm A}>1$ cases (LV99). The former, the case of turbulence driven with a sub-Alfv{\'e}nic velocity, shows weak turbulence ranging from $L_{\rm inj}$ to the transition scale 
\begin{equation}
l_{\rm trans}=L_{\rm inj}M_{\rm A}^2. 
\end{equation}
Here, the perturbations of magnetic fields are quasi-2D and perpendicular to the direction of magnetic fields (LV99, \citealt{Galtier00}), which indicates that the gradients of magnetic fields are also as perpendicular to the magnetic fields as those of GS95. At scales less than $l_{\rm trans}$, the strong sub-Alfv{\'e}nic turbulence is present, the eddies of which are more elongated along the local magnetic field than those in the original GS95. Over the inertial range of $[l_{\rm diss}, l_{\rm trans}]$, where $l_{\rm diss}$ is a dissipation scale, the relation between the eddy length scale $l_\|$ and its transversal extent $l_\bot$ is  
\begin{equation}
l_{\|}\approx L_{\rm inj}\left(\frac{{ l_\perp}}{L_{\rm inj}}\right)^{2/3} M_{\rm A}^{-4/3}. \label{eq:aniso1}
\end{equation}
When $M_A=1$, Equation (\ref{eq:aniso1}) goes back to the original GS95 relation (Equation (\ref{eq:aniso})). 

The latter corresponds to super-Alfv{\'e}nic turbulence, $V_{\rm L}> {V}_{{\rm{A}}}$, i.e., ${M}_{{\rm{A}}}>1$. As for a limiting case of ${M}_{{\rm{A}}}\gg 1$, the turbulence when approaching to the injection scale possesses an essentially hydrodynamic Kolmogorov behavior, i.e., $v_l=V_{\rm L}(l/L_{\rm inj})^{1/3}$, due to 
the marginal influence of the weak magnetic fields on MHD turbulence. When turbulent velocity equals the Alfv{\'e}nic velocity, i.e., $v_{l}=V_{\rm A}$, the hydrodynamic-like characteristic of the turbulence cascade presents a transition at the scale
\begin{equation}
\label{SupLA}
l_{\rm A}=L_{\rm inj}M_{\rm A}^{-3}.
\end{equation}
No correlation happens between the gradient directions of magnetic fields and the directions of magnetic fields at the larger scale than $l_{\rm A}$. It is easy to understand that no alignment of eddies and magnetic fields is expected when hydrodynamic-like motion dominates. On the other hand, significant correlation appears between them at the smaller scale than $l_{\rm A}$ when magnetic fields dominate again.

The symmetry between the magnetic field and the eddy velocity makes their gradients perpendicular to the directions of the local magnetic field. This fundamental fact inspires the development of velocity and synchrotron gradient techniques for tracing magnetic fields. According to the relation of $v_{l}/l_{\perp}\propto l_{\perp}^{-2/3}$ derived from the previous description, it can be inferred that the smallest eddies correspond to the largest gradients that are perpendicular to local magnetic fields and determine the local magnetic field directions. In this paper, we rotate 90 degrees for the magnetic field gradients to identify the directions of potential magnetic fields.  

\subsection{Synchrotron emission statistical diagnostics}
We consider that non-thermal relativistic electrons have a power-law energy distribution of $N_e\propto\gamma^{-p}$, where $p$ and $\gamma$ are the spectral index and energy of electrons, respectively, and combine the formulae commonly used in the study of synchrotron polarization emission (see the Appendix of \citealt{Waelkens09} and LP16). The well-known Stokes parameters $I$, $Q$, $U$ and $V$ are usually used to depict the synchrotron emission, characterizing different behaviors of synchrotron radiative processes. Our work focuses mainly on statistical measurements related to the first three parameters. 

As for linearly polarized synchrotron emission as mentioned in Section \ref{intro}, polarization information is usually described by polarized angle of $\theta\equiv {1\over 2} {\rm arctan}(U/Q)$ and polarization degree of $\Pi=\sqrt{Q^2+U^2}/I$. However, these quantities cannot be maintained under the condition of arbitrary translation and rotation in the complex $Q$-$U$ plane. A remarkable progress with insight into physical properties of the ISM has been made by \cite{Gaensler11} formulating synchrotron polarization fluctuations. Specifically, they proposed new physical quantities: the magnitude of the gradients of the polarization vector 
\begin{equation}
P_{\rm v}=|\nabla  \textit{\textbf{P}}| =\sqrt{ \left({\partial Q\over \partial x}\right)^2 +  \left({\partial U\over \partial x}\right)^2  +  \left({\partial Q\over \partial y}\right)^2  +  \left({\partial U\over \partial y}\right)^2  }\label{eq:pv}
\end{equation}
and the gradient directions of the polarization vector
\begin{widetext}
\begin{equation}
\begin{aligned}
{\rm arg}(\nabla \textit{\textbf{P}})=  {\rm arctan}\left[{\rm sign}\left({\partial Q\over \partial x}{\partial Q\over \partial y} +{\partial U\over \partial x}{\partial U\over \partial y} \right) 
\sqrt{\left({\partial Q\over \partial y}\right)^2 +\left({\partial U\over \partial y}\right)^2}\Bigg /\sqrt{\left({\partial Q\over \partial x}\right)^2 +\left({\partial U\over \partial x}\right)^2}  \right]. \label{eq:pv_ang}
\end{aligned}
\end{equation}
\end{widetext}

Taking the translational and rotational invariance into account, \cite{Herron18a} developed a mathematical formalism to derive extra invariant quantities. The formulae relevant to our current work go as follows: 
\begin{widetext}
\begin{align}
P_{\rm vg} = & |\nabla  \textit{\textbf{P}}| = \Biggl[ \frac{1}{2} \biggl( \biggl(\frac{\partial Q}{\partial x} \biggr)^2 + \biggl(\frac{\partial U}{\partial x} \biggr)^2 + \biggl(\frac{\partial Q}{\partial y} \biggr)^2 + \biggl(\frac{\partial U}{\partial y} \biggr)^2  \biggr) + \nonumber \\ 
 & \frac{1}{2} \sqrt{\biggl( \biggl(\frac{\partial Q}{\partial x} \biggr)^2 + \biggl(\frac{\partial U}{\partial x} \biggr)^2 + \biggl(\frac{\partial Q}{\partial y} \biggr)^2 + \biggl(\frac{\partial U}{\partial y} \biggr)^2 \biggr)^2 - 4 \biggl( \frac{\partial Q}{\partial x} \frac{\partial U}{\partial y} - \frac{\partial Q}{\partial y} \frac{\partial U}{\partial x} \biggr)^2  } \Biggr]^{1/2}  \label{eq:pvg}
\end{align}
\end{widetext}
for the generalization of the polarization gradient;
\begin{equation}
P_{\rm rad} = \sqrt{ \frac{ \bigl( Q \frac{\partial Q}{\partial x} + U \frac{\partial U}{\partial x} \bigr)^2 + \bigl( Q \frac{\partial Q}{\partial y} + U \frac{\partial U}{\partial y} \bigr)^2 }{ Q^2 + U^2 } }  \label{eq:prad}
\end{equation}
for the maximum of the radial component of the polarization directional derivative and 
\begin{equation}
P_{\rm tang}= \sqrt{ \frac{ \bigl( Q \frac{\partial U}{\partial x} - U \frac{\partial Q}{\partial x} \bigr)^2 + \bigl( Q \frac{\partial U}{\partial y} - U \frac{\partial Q}{\partial y} \bigr)^2 }{ Q^2 + U^2 } }  \label{eq:ptang}
\end{equation}
for the maximum of the tangential component. Equations (\ref{eq:prad}) and (\ref{eq:ptang}) describe imaging changes in polarization intensity and polarization angle, respectively.

When it comes to Faraday rotation effect, the Stokes parameters $I$, $Q$ and $U$ are obtained in our numerical treatment by 
\begin{equation}
\textbf{\textit{P}}(\textbf{\textit{X}},\lambda^2)=\int^{L}_{0}dzP_i(\textbf{\textit{X}},z){e}^{2i\lambda^2\phi(\textbf{\textit{X}},z)}, \label{PEq}
\end{equation}
where $L$ is the emitting region extent along the LOS ($z$-axis) direction. $P_i(\textbf{\textit{X}},z)$ indicates the density of the intrinsic polarized intensity, being wavelength independent. Such simplification would have no impact on our statistic results (\citealt{Zhang18}). The Faraday rotation measure is illustrated by 
\begin{equation}
\phi(\textbf{\textit{X}},z)={e^3\over2\pi m_{\rm e}^2c^4}\int^z_{0}dz'n_{\rm e}(\textbf{\textit{X}},z')B_z(\textbf{\textit{X}},z'), \label{RM}
\end{equation}
where $n_{\rm e}$ and $B_{ z}$ refer to thermal electron density and parallel component of magnetic fields, respectively. 

\subsection{Gradient determination}\label{GMT}
The recipe of sub-block averaging (\citealt{GL17}, hereafter GL17) is used to calculate the gradients of synchrotron emission diagnostics and potential magnetic fields. In each subregion of interest, we take the average of the gradient angles to get an optimal direction characterized by the peak of the Gaussian fitting. Accordingly, the corresponding directions of rotated $90^\circ$ gradients would identify the directions of mean magnetic fields. In practice, GL17 defined a reduction factor to measure the correspondence between velocity gradients and the magnetic field, analogous to the Rayleigh reduction factor in dust alignment theory suggested by \cite{Greenberg68}. This factor is what is so called the AM given by 
\begin{equation}
AM=2\langle\cos^2\varphi\rangle-1, \label{eq:AM}
\end{equation}
which determines an alignment between the rotated $90^\circ$ gradients and magnetic field directions, with $\varphi$ as the angle between the gradients of 
statistical diagnostics and the magnetic fields. Compared with the early Rayleigh reduction factor, the difference lies in that the AM is for a 2D distribution instead of a 3D one. A perfect alignment can be achieved with $AM =\pm1$, indicating the gradients perpendicular or parallel to the projected magnetic field, while no alignment
occurs with $AM=0$.

\section{Numerical simulations}\label{nums}
Data cubes with numerical resolution of $792^3$ in this work are generated by a single-fluid, operator-split, staggered-grid MHD Eulerian code ZEUS-MP/3D (\citealt{Hayes06}). A periodic boundary condition combining with a solenoidal turbulence injection is assumed to simulate a three-dimensional and isothermal turbulent medium. We explore a wide range of Alfv{\'e}nic Mach numbers in the low $\beta<1$ turbulence scenario. Specific numerical parameters are listed in Table \ref{table:simdata}.

\begin{table}
 \caption {Data cubes of numerical resolution of $792^3$ in this paper are generated by using open source ZEUS-MP code. We mark $\delta B_{\rm rms}$ as the root mean square random magnetic field and $\left<B\right>$ as the regular magnetic fields along the $x$-axis. }
 \centering
 \begin{tabular}{c c c cc}%
 \hline
Model & $M_{\rm A}$ & $\beta=2M_{\rm A}^2/M_{\rm s}^2$ & $\delta B_{\rm rm s} / \langle B \rangle$ & Description\\ \hline 
run1 & 0.22  & 0.002 &  0.15 & Sub-Alfv{\'e}nic  \\
run2& 0.42 & 0.011  & 0.36  &  Sub-Alfv{\'e}nic  \\
run3 & 0.61 & 0.022  & 0.47 & Sub-Alfv{\'e}nic   \\
run4 & 0.82 & 0.042  & 0.63 & Sub-Alfv{\'e}nic   \\
run5 & 1.01 &  0.065  & 0.76 & Trans-Alfv{\'e}nic  \\
run6 & 1.19 & 0.089  & 0.87 & Super-Alfv{\'e}nic    \\ 
run7 & 1.38 & 0.118   & 1.02 & Super-Alfv{\'e}nic  \\ 
run8 & 1.55 & 0.155   & 1.12 & Super-Alfv{\'e}nic   \\
run9 & 1.67 & 0.184   & 1.25 & Super-Alfv{\'e}nic  \\
run10 & 1.71 & 0.201   & 1.39 & Super-Alfv{\'e}nic \\ \hline 
\end{tabular}
\label{table:simdata}
\end{table}

\subsection{Structure of synchrotron multifarious diagnostics}

\begin{figure*}[t]
\centering
\includegraphics[width=1.0\textwidth]{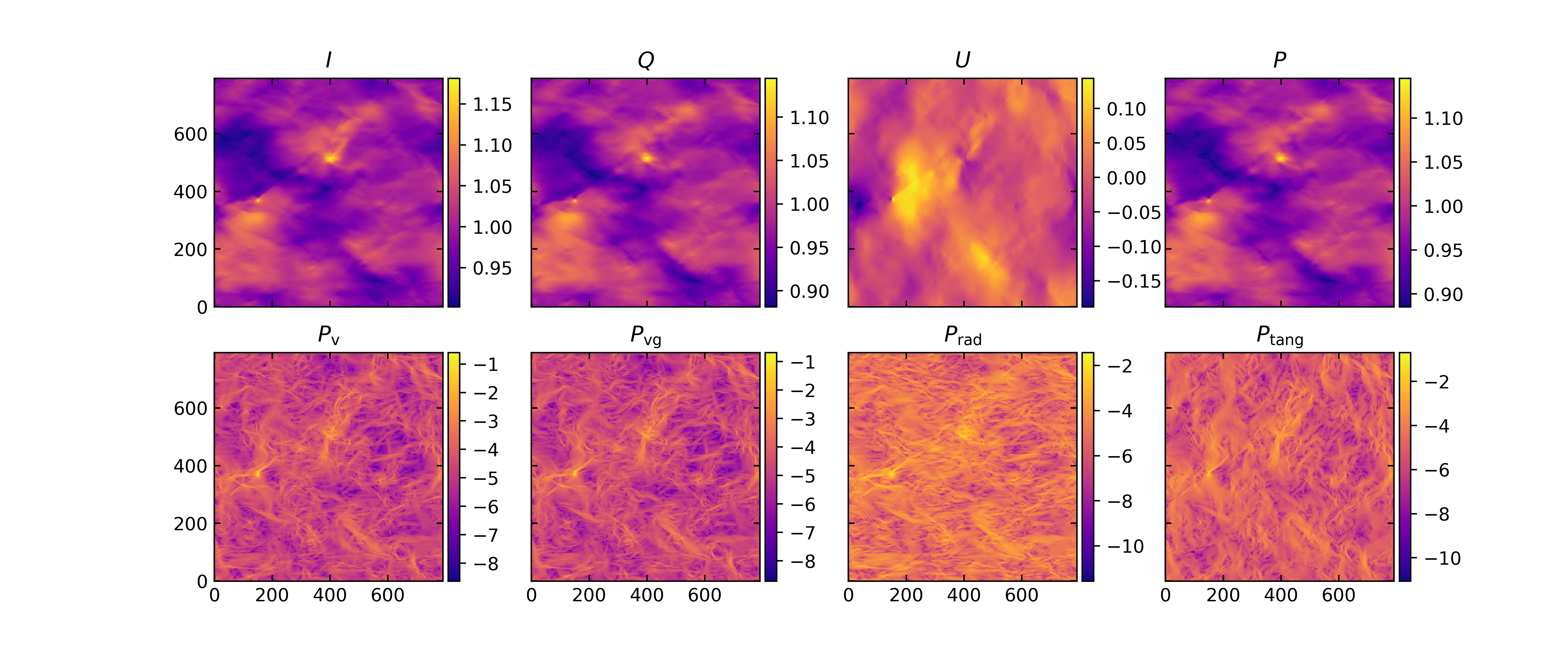}
\caption{Images of different diagnostics of synchrotron emission with no Faraday rotation. Maps plotted via the run1 of Table \ref{table:simdata}, cover Stokes parameters of $I$, $Q$ and $U$, polarization intensity of $P$, amplitude of polarization vector gradient of $P_{\rm v}$, amplitude of generalized polarization vector gradient of $P_{\rm vg}$ as well as radial of $P_{\rm rad}$ and tangential of $P_{\rm tang}$ components of the polarization directional derivative. Results are displayed in units of mean synchrotron intensities but on logarithmic scales in the lower panels. } \label{fig:map_no_FR}
\end{figure*}

Based on run1 of Table \ref{table:simdata} for sub-Alfv{\'e}nic turbulence, we present in Figure \ref{fig:map_no_FR} maps of multifarious diagnostics of synchrotron emission with Faraday rotation effect ignored. All the images are normalized by mean synchrotron intensity, and the colorbars in upper and lower panels are shown on normal and logarithmic scales, respectively, in order to observe detailed features. 

In the upper panels of Figure \ref{fig:map_no_FR}, it is shown that maps of $I$, $Q$ and $P$ have nearly similar structures extending horizontally but different amplitudes. These phenomena can be illustrated by the following relations:\footnote{In this paper, we assume the spectral index of relativistic electrons to be $p=3$. No changes could happen to our statistical results with alternative $p$ values assigned, as is strengthened previously (\citealt{Zhang16}).} $I\propto B_{\perp}^{p-3\over 4} (B_x^2+B_y^2)$, $Q\propto -B_{\perp}^{p-3\over 4} (B_x^2-B_y^2)$ and $U\propto -B_{\perp}^{p-3\over 4} (B_xB_y)$, where $B_{\perp}=\sqrt{B_x^2+B_y^2}$ is the perpendicular component in the plane of the sky of 3D magnetic fields. Similarities of extended structures and differences of amplitudes between maps of $I$ and $Q$ can be caused by the dominated role of non-zero $\left<B_x\right>$ in synchrotron emission fluctuations. As for the structure in the map of $P$, it reflects the intrinsic polarization information of an emitting source without the influence of Faraday rotation. The structure in the map of $P$ ($=\sqrt{Q^2+U^2}$) arising from stronger fluctuations of $Q$ than those of $U$ are similar to that of $I$. Roughly speaking, the structures extending horizontally along the $x$-axis in the maps of $I$, $Q$ and $P$ can disclose the directions of mean magnetic fields. As can be seen, the structure in the map of $U$ presents to be almost isotropic owing to the non-linear combination of $B_xB_y$.

In the lower panels of Figure \ref{fig:map_no_FR} are imaged the amplitudes of polarization vector gradient (labeled as $P_{\rm v}$, see Equation (\ref{eq:pv})), of generalized polarization vector gradient ($P_{\rm vg}$ in Equation (\ref{eq:pvg})) as well as of radial ($P_{\rm rad}$ in Equation (\ref{eq:prad})) and tangential ($P_{\rm tang}$ in Equation (\ref{eq:ptang})) components of the polarization directional derivative. More detailed filamentary features emerge from the imaged amplitudes above than in the upper panels. The fact that the more anisotropic the structure is, the larger the AM is and the better the directions of magnetic fields is traced can be explored below.

\subsection{Alignment measure without Faraday rotation}

\begin{figure}[t]
\centering
\includegraphics[width=0.5\textwidth]{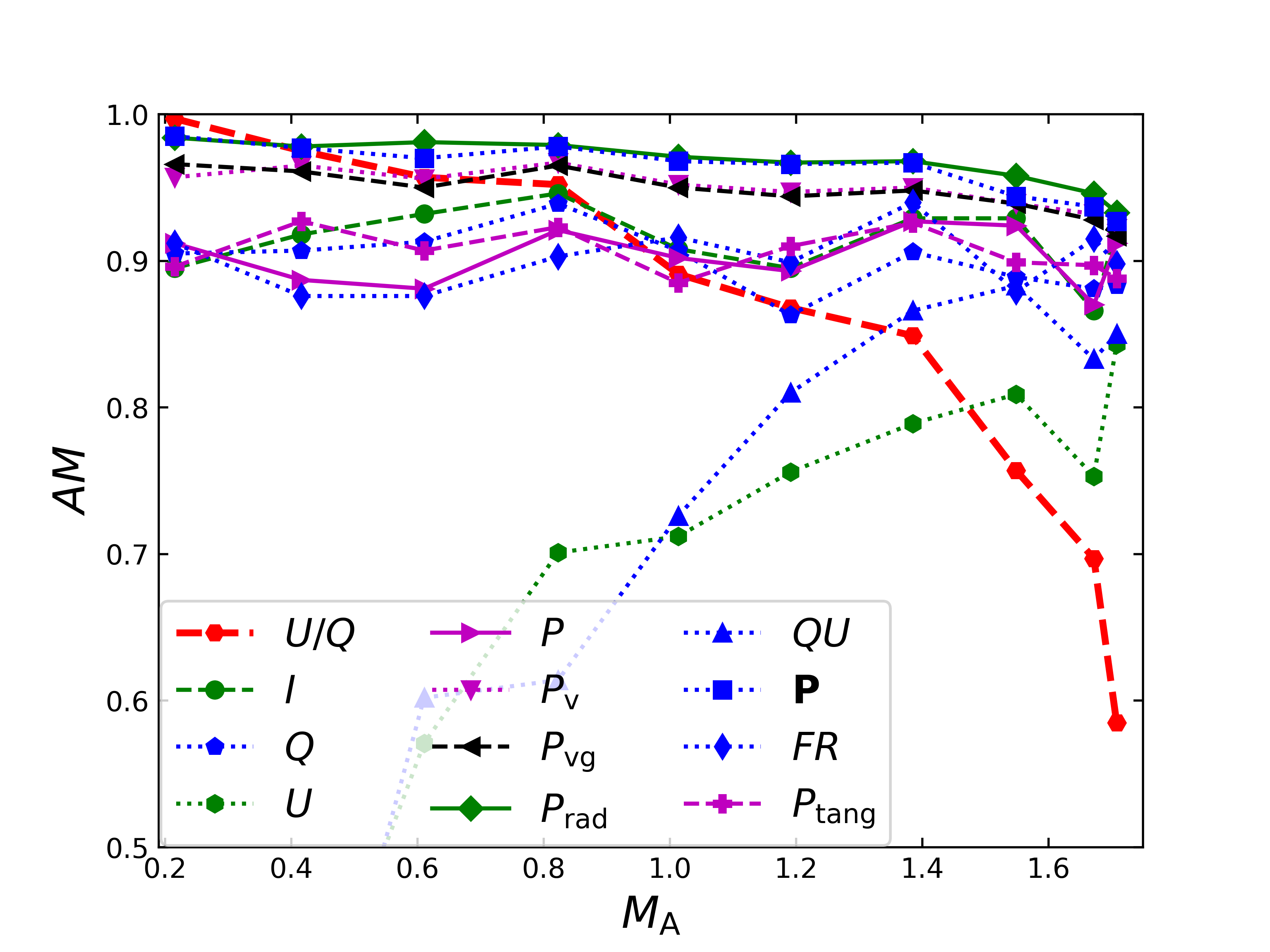}
\caption{AM, a function of Alfv{\'e}nic Mach number, between the projected magnetic fields and rotated gradients for multifarious synchrotron diagnostics in the case of negligible Faraday depolarization. The legend of \textit{FR}, $U/Q$ and $QU$ demonstrate the following alignment respectively: of Faraday rotation gradient, of traditional polarization vector (rotated 90 degrees), of cross-intensities gradients of both $Q$ and $U$, to projected magnetic fields. Other legends have common meanings. A Gaussian kernel of $\sigma=2$ pixels is used to smooth noise-like structures that affect the measurement of both the gradients and traditional polarization vector.
} \label{fig:AM_no_FR}
\end{figure}

\begin{table*}[t]
\caption {Descriptions of diagnostic parameters used in this paper. }
 \centering
 \begin{tabular}{c c c }%
 \hline 
Symbol  & Description &  Reference \\ \hline \hline
$I$ &  Stokes parameter &  Section 2.2\\
$Q$ &  Stokes parameter &  Section 1\\
$U$ &  Stokes parameter &  Section 1\\
$P$ &  Polarization intensity or amplitude of polarization intensity ($\sqrt{Q^2+U^2}$) & Section 3.1\\
$\textit{\textbf{P}}$ & Complex polarization vector & Section 1\\
$QU$ &  Cross-intensities defined as the product of Stokes $Q$ and $U$ &  Section 3.2\\
$U/Q$ &  Traditional polarization vector measure by $\theta\equiv {1\over 2} {\rm arctan}(U/Q)$  &  Section 3.2\\
 $P_{\rm v}$ & Magnitude of polarization vector gradients & Equation (5)\\
 $P_{\rm vg}$ & Magnitude of generalized polarization gradients & Equation (7)\\
 $P_{\rm rad}$ & Maximum of the radial component of the polarization directional derivative & Equation (8)\\
 $P_{\rm tang}$ & Maximum of the tangential component of the polarization directional derivative & Equation (9)\\
  $\phi$ & Faraday rotation (\textit{FR}) measure & Equation (11)\\ \hline \hline
\end{tabular}
 \label{table:para}
\end{table*}

In the case of negligible Faraday rotation effect, polarized synchrotron radiation is calculated at high frequency of $10^3$ GHz. AMs of rotated $90^\circ$ gradients of multifarious synchrotron diagnostics vs. projected magnetic field directions, as a function of Alfv{\'e}nic Mach number, are shown in Figure \ref{fig:AM_no_FR}. Gradients of $I$, $Q$, $FR$, $P$, $P_{\rm v}$, $P_{\rm vg}$, $P_{\rm rad}$ and $\textit{\textbf{P}}$ (see Table \ref{table:para} for their respective meanings) are found to be able to trace the directions of the potential mean magnetic fields, where gradients of all scalars are calculated following the improved procedure of velocity centroid gradients (GL17) and gradients of $\textit{\textbf{P}}$ vector are performed through Equation (\ref{eq:pv_ang}). In Figure \ref{fig:AM_no_FR} and below, we use $FR$ for gradients of Faraday rotation measure labeled as $\phi$ in Equation (\ref{RM}). What is worth mentioning here is that other synchrotron diagnostics do not include the Faraday depolarization effect in this section. In other words, all measures explored in this section are frequency independent. In comparison with traditional polarization method (i.e., $U/Q$), all the above diagnostics can trace the directions of projected magnetic fields in a reliable manner, even in the super-Alfv{\'e}nic regime. 

As is shown in Figure \ref{fig:AM_no_FR}, gradients of $U$ and $QU$ are not credible for magnetic field tracing within the sub-Alfv{\'e}nic range, where $QU$ stands for gradients of cross-intensities of $Q$ and $U$, i.e., the product of $Q$ and $U$. It is the coordinate system selection that causes the reduced AM values for $U$ and $QU$, which will be improved in the case of strong Faraday rotation (see Section \ref{MultiFM}) or of rotated mean magnetic field orientations (Section \ref{IMMFD}). The need for rotationally-invariant quantities should be emphasized as an important part of this work and of \cite{Gaensler11} and \cite{Herron18a}. With the increase of $M_{\rm A}$, all the effective AMs experience slight reduction, which is in agreement with simulations presented by \cite{Zhang19} in the case of low resolution numerical simulations. Nevertheless, it is more challenging for magnetic field tracing in larger $M_{\rm A}$ regime.

\subsection{Alignment measure with Faraday rotation}
\subsubsection{Structure of Faraday rotation measure}
In consideration of the Faraday rotation effect, it is extremely complicated in the study of polarization radiation to unveil the properties of magnetic turbulence, despite nothing but a mathematical exponential factor added in Equation (\ref{PEq}). Thus, we focus in our work  on the scenario of the spatially coincident polarized synchrotron emission and Faraday rotation regions of the ISM, with the assumption that the emission region is extended to be 1 kpc along the LOS, and thermal electron density and magnetic field strength are set as $n_{\rm e}\sim0.01\ \rm cm^{-3}$ and $B_{ z}\sim 1.3\ \mu\rm G$, respectively. 

\begin{figure*}[t]
\centering
\includegraphics[width=1.0\textwidth]{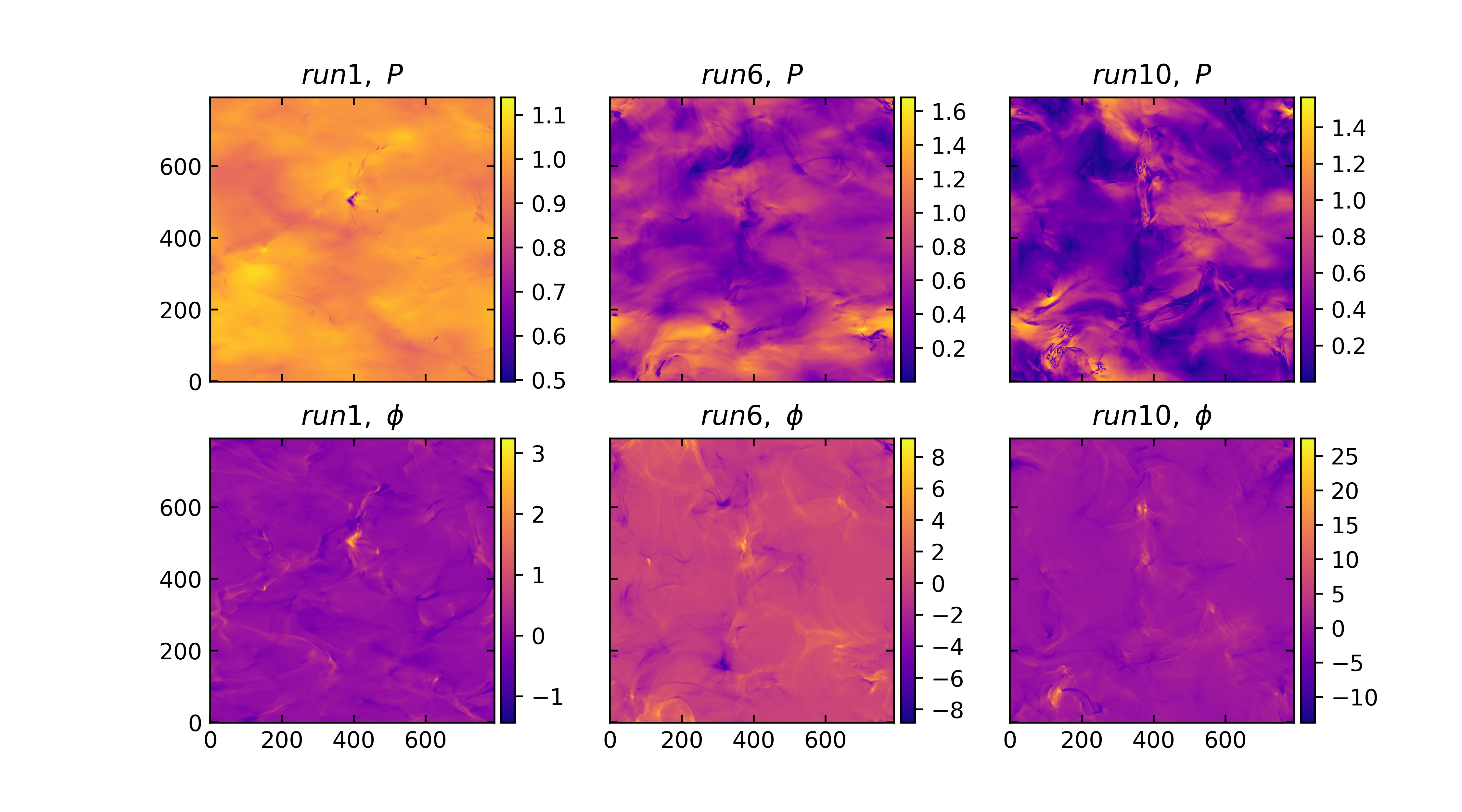}
\caption{Polarized synchrotron intensities (upper panels, in units of mean synchrotron intensity) and Faraday rotation measure (lower panels, in units of rad $\rm m^{-2}$) calculated via run1 (left column), run6 (middle column) and run10 (right column) listed in Table \ref{table:simdata}, respectively. 
} \label{fig:FRdepth}
\end{figure*}

Figure \ref{fig:FRdepth} shows 2D structure of Faraday rotation measure integrated along the LOS in the lower panels, in comparison with that of polarized intensities in the upper panels. Noticeably, amplitude values of Faraday rotation measure in the sub-Alfv\'enic turbulence regime (left lower panel) are seen to be smaller than those in the super-Alfv\'enic regime (see middle and right lower panels). The larger $\phi$ value is, the stronger Faraday rotation effect would become.
A positive value indicates the direction of magnetic fields along the LOS to the observer, while a negative value the direction of magnetic fields along the LOS away from the observer. As for polarization intensity in the upper panels, we find a phenomenon that in the local region of the image, the stronger the Faraday rotation is, the weaker the polarization intensity becomes. 

\subsubsection{Multifrequency measure from different turbulence regimes}\label{MultiFM}
\begin{figure}[t]
\centering
\includegraphics[width=0.5\textwidth]{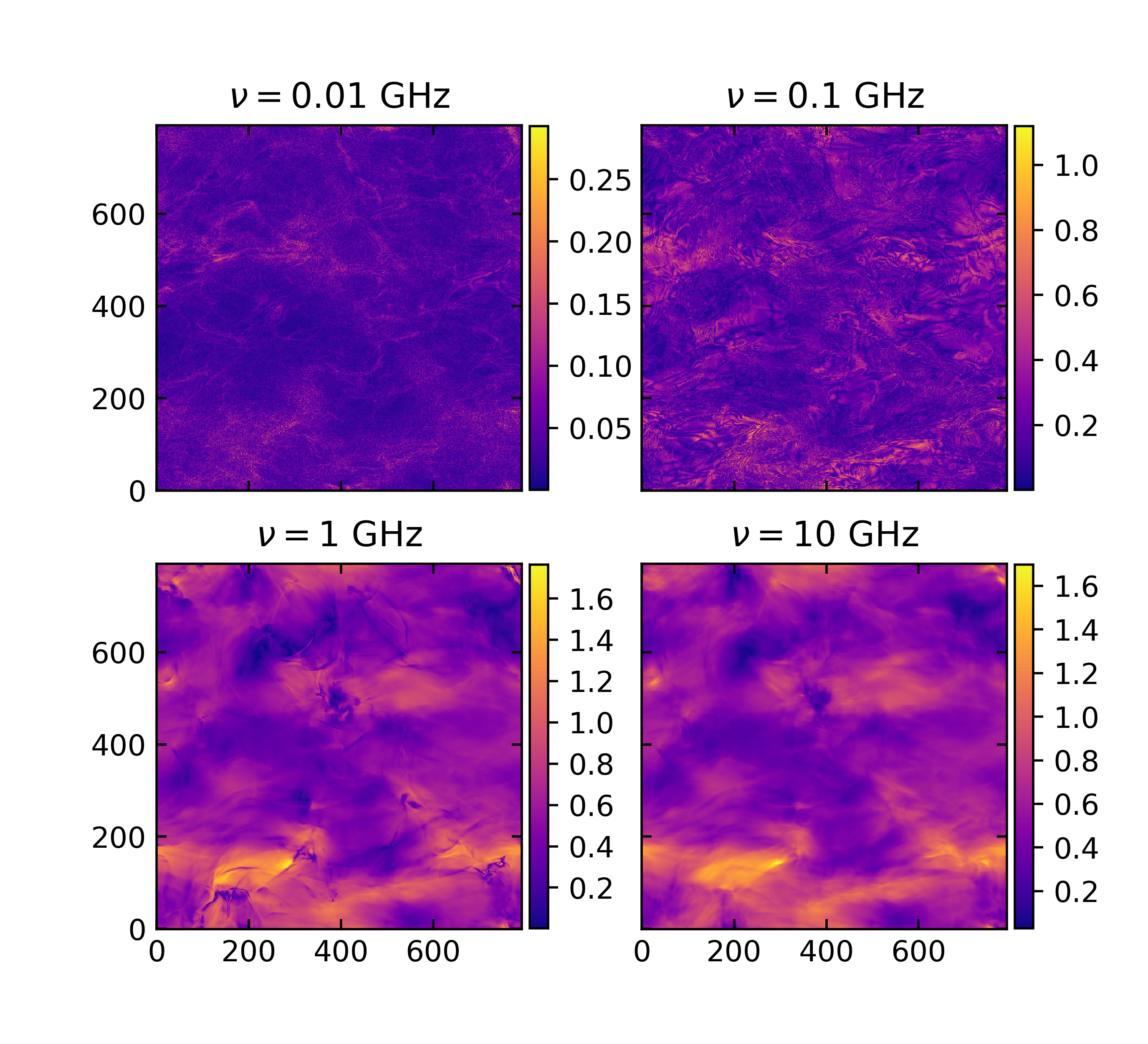}
\caption{Polarized synchrotron intensities in units of mean synchrotron intensity at different frequencies, on the basis of run7 listed in Table \ref{table:simdata}. 
} \label{fig:Polnoise}
\end{figure}

Investigation into the effect the frequency changes have on the structure of the polarization intensity image comes before an alignment analysis. Based on run7 in Table \ref{table:simdata}, four images of polarized intensities are plotted in Figure \ref{fig:Polnoise} at frequencies $\nu=0.01$, 0.1, 1 and 10 GHz, respectively. As a result, we find amplitudes of polarization intensities decrease with the frequency (appearance of positive correlation between frequency and amplitudes) with more small-scale noise-like structures emerging in their maps. The reason why is that polarized signals suffer from a more significant Faraday rotation depolarization at low frequencies. To eliminate the influence of noise structure on the gradient measurement technique, Gaussian filtering technique with a kernel $\sigma=2$ is used in this paper.

\begin{figure*}[t]
\centering
\includegraphics[width=0.45\textwidth,bb=5 50.5 420 360]{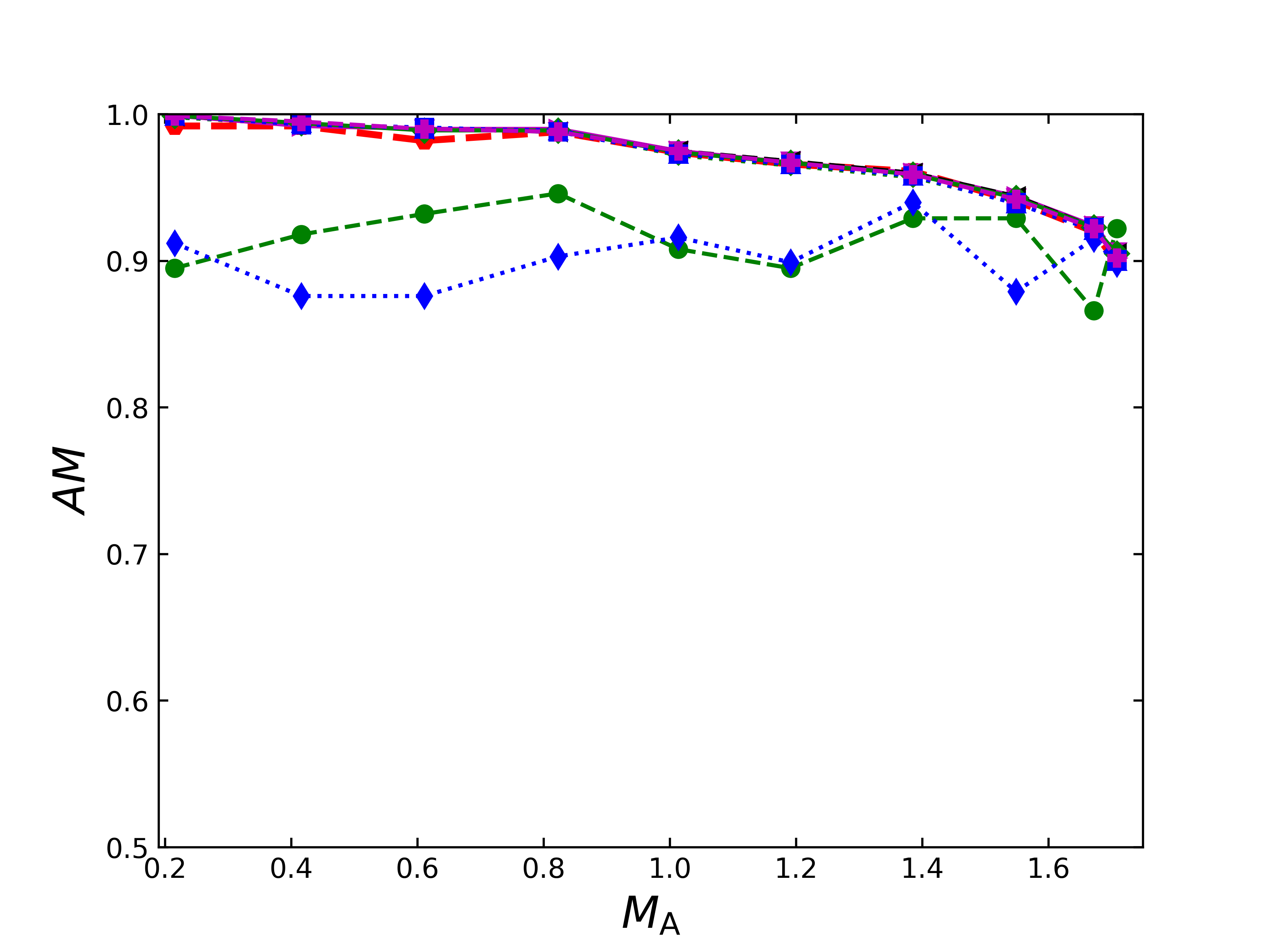}
\includegraphics[width=0.45\textwidth,bb=5 50.5 420 360]{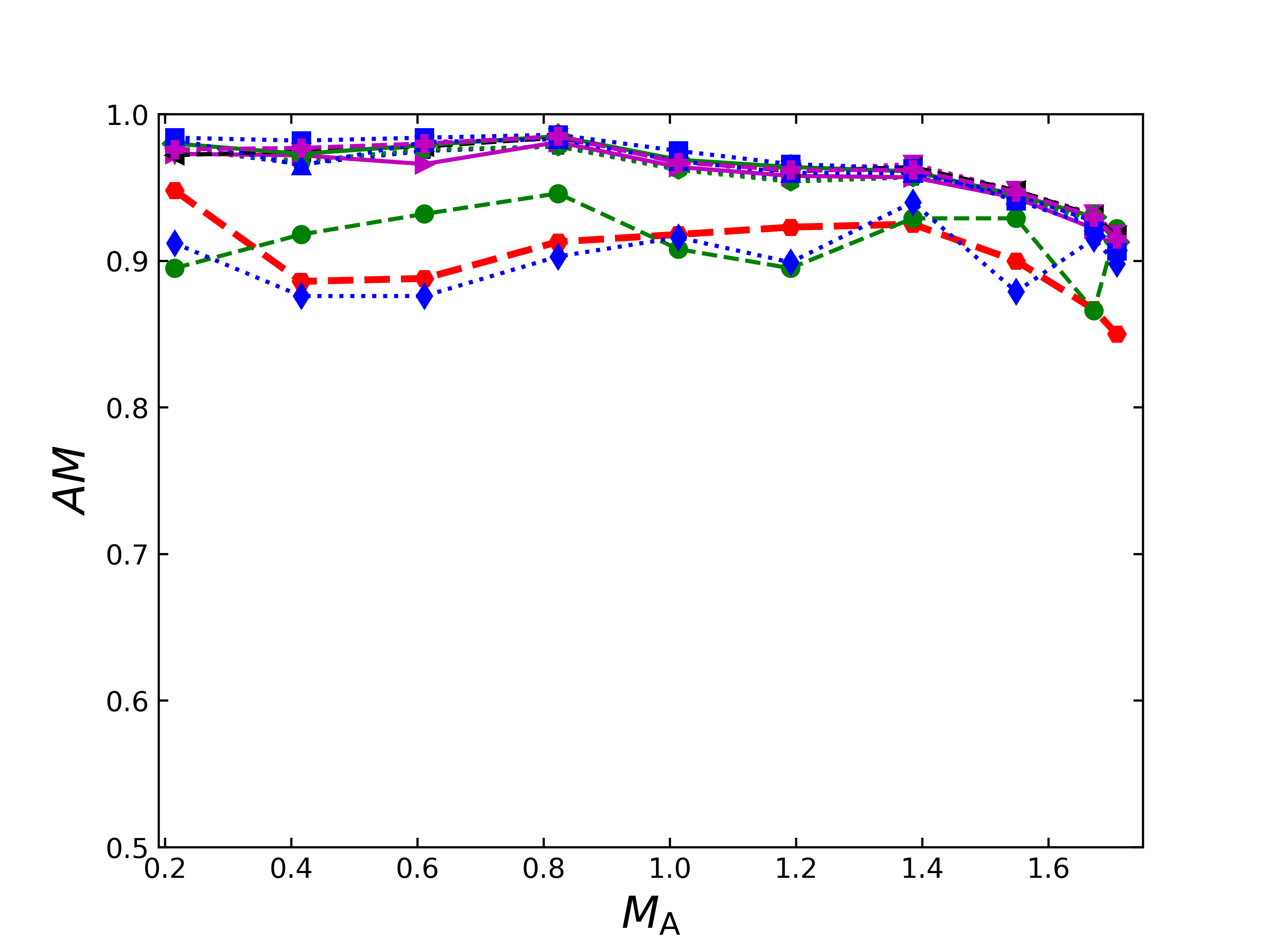}
\includegraphics[width=0.45\textwidth,bb=5 10 420 360]{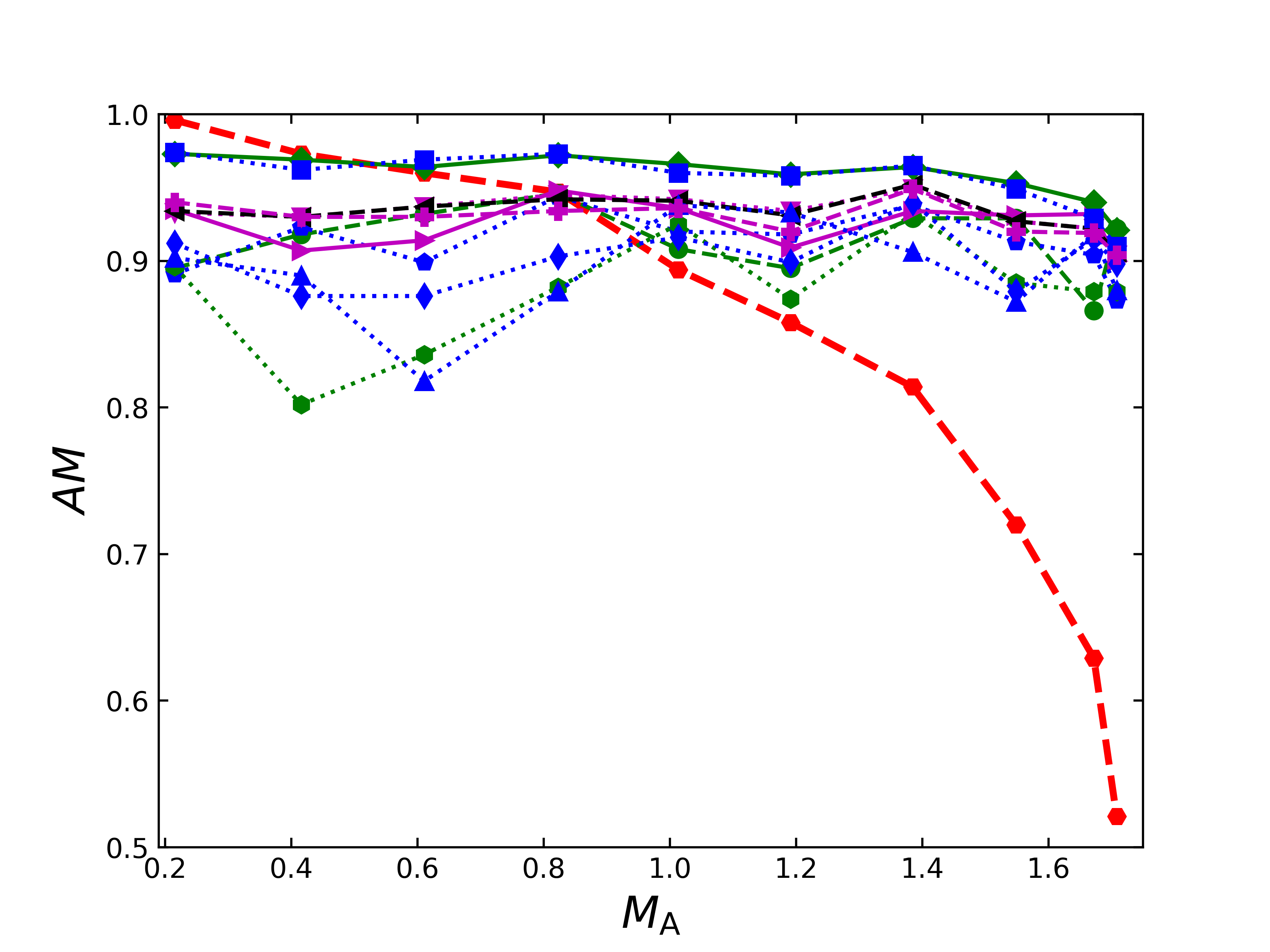}
\includegraphics[width=0.45\textwidth,bb=5 10 420 360]{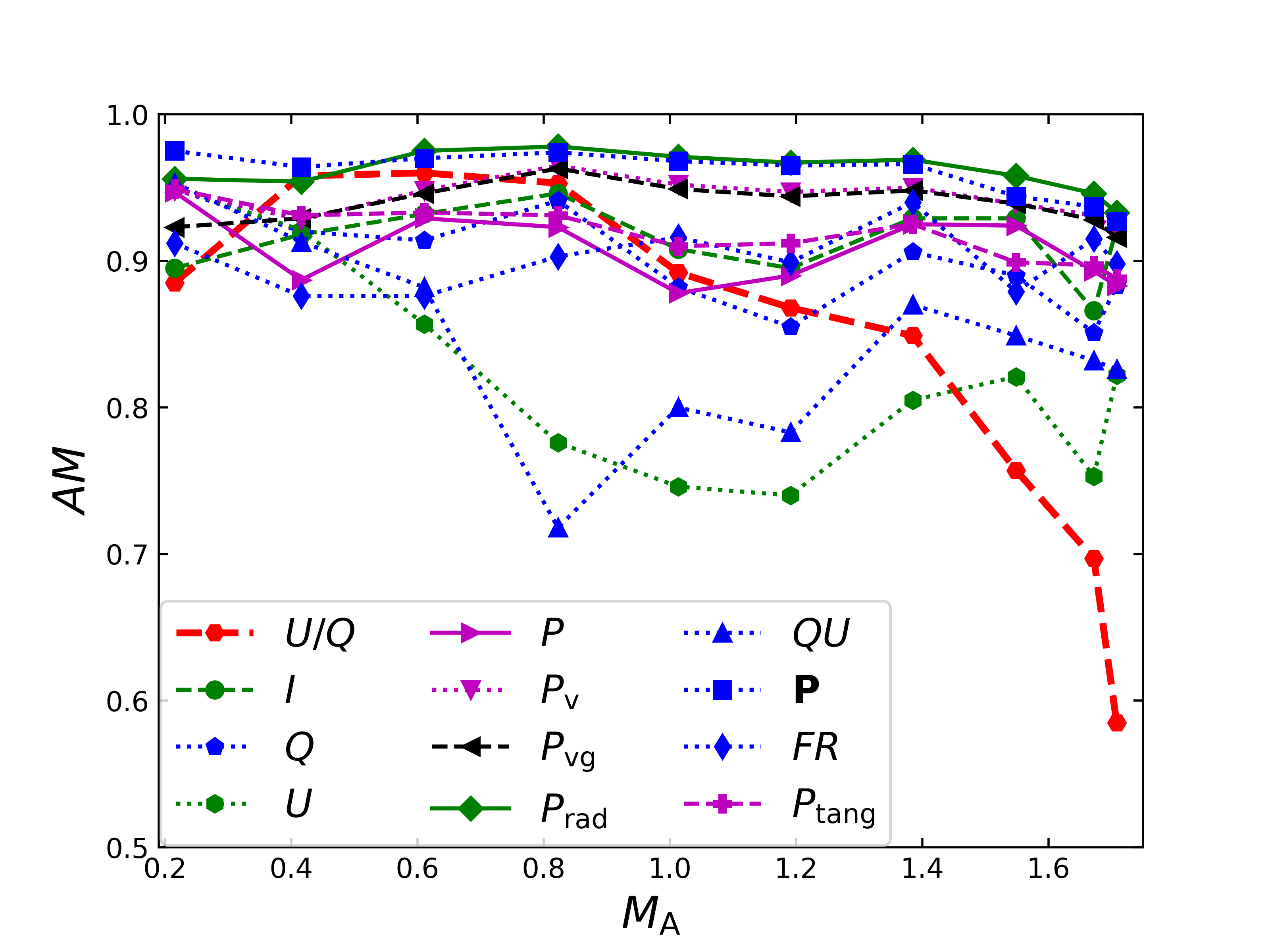}
\caption{AM, as a function of Alfv{\'e}nic Mach number, between rotated $90^\circ$ gradients of multifarious synchrotron diagnostics and projected magnetic field directions at frequencies $\nu=0.01$ (left upper), 0.1 (right upper), 1 (left lower) and 10 GHz (right lower). Other descriptions are the same as those in Figure \ref{fig:AM_no_FR}. 
 } \label{fig:AM_with_FR} 
\end{figure*}

Figure \ref{fig:AM_with_FR} shows the influence of turbulence magnetization on AM at frequencies $\nu=0.01$ (left upper), 0.1 (right upper), 1 (left lower) and 10 GHz (right lower). The resulting AM demonstrates that gradient measure of multifarious diagnostics is successful in determining directions of projected magnetic fields even in the rather low frequency and strong Faraday rotation regime. It can be seen from this figure that gradients of $U$ and $QU$ in the case of Faraday rotation are shown to work better than those in weak or no Faraday depolarization (see Figure \ref{fig:AM_no_FR}). We claim that low-frequency strong Faraday rotation breaks the alignment dependence on the coordinate system, making $Q$ and $U$ effectively equivalent in these simulations. Of multifarious diagnostics, $\textit{\textbf{P}}$ and $P_{\rm rad}$ appear to be the most significant measurements. As $M_{\rm A}$ increases, AM values slightly decrease resulting from the presence of the weaker magnetic field and large Faraday rotation measure (see Figure \ref{fig:FRdepth}) in the super-Alfv{\'e}nic turbulence. In a word, using synergies of multifarious diagnostics can surely enhance the reliability to trace turbulent magnetic fields.

\begin{figure*}[t]
\centering
\includegraphics[width=0.45\textwidth]{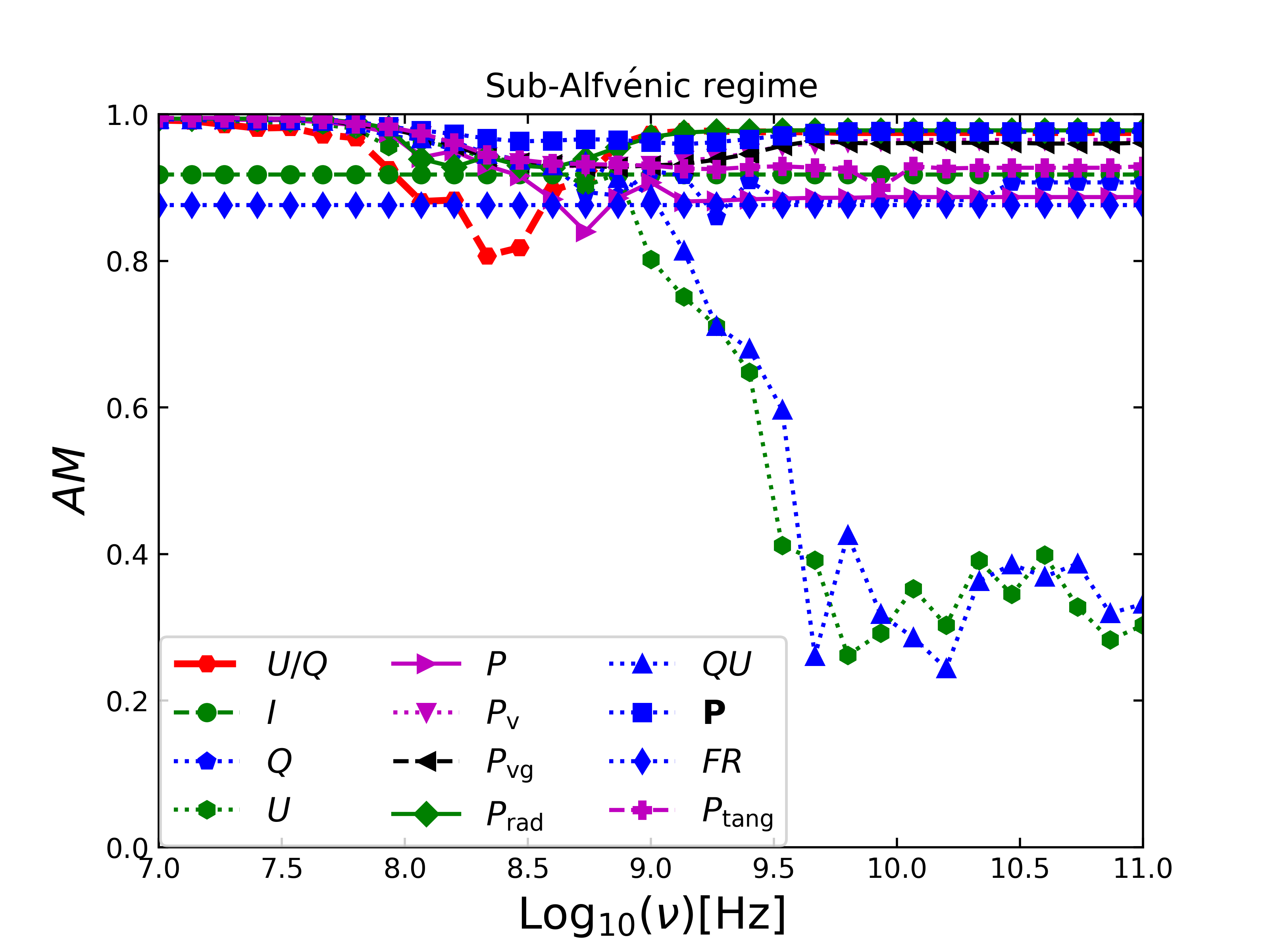}
\includegraphics[width=0.45\textwidth]{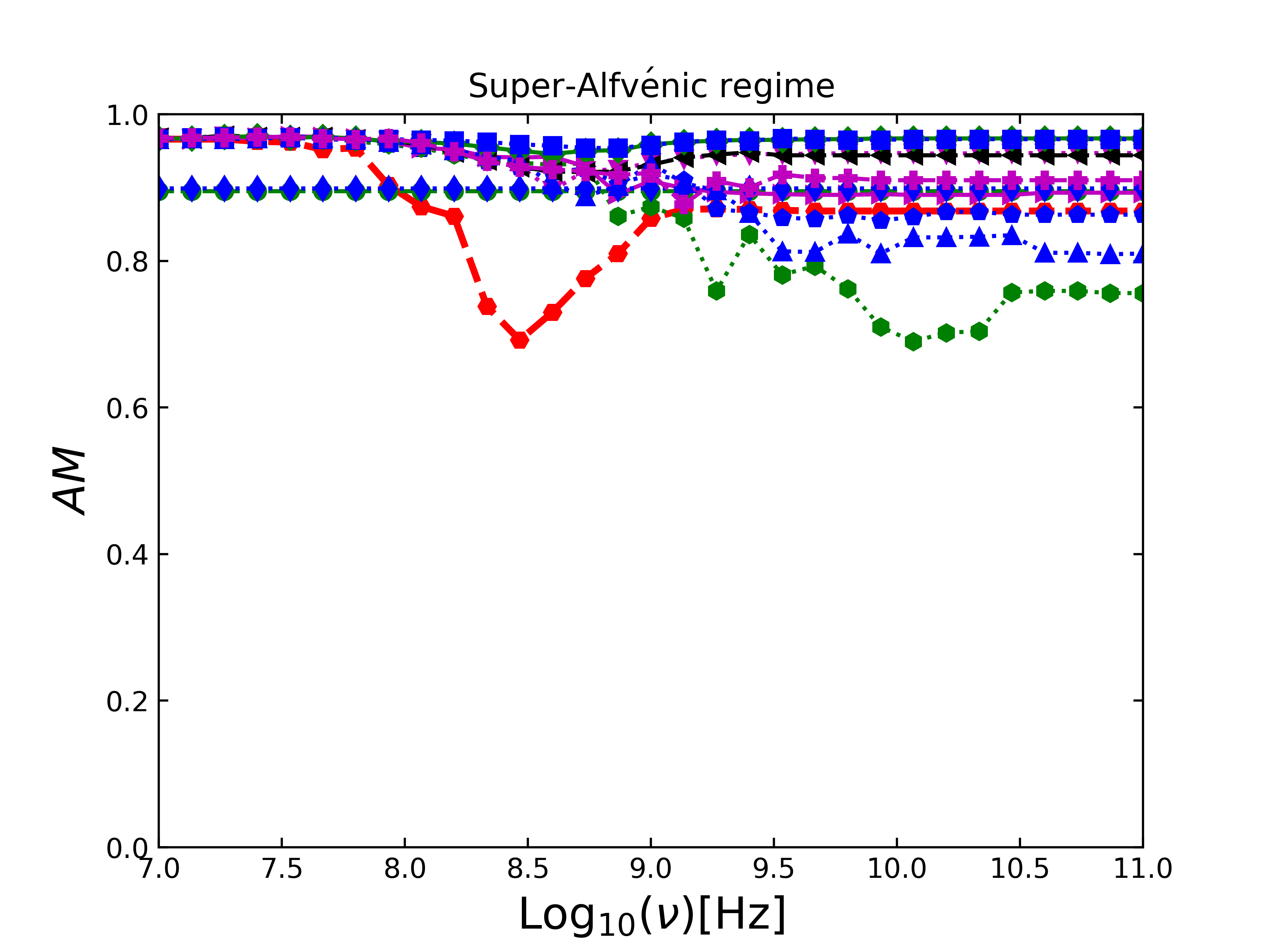}
\caption{AM, as a function of frequency, between directions of mean magnetic fields and directions predicted by multifarious diagnostics in sub- and super-Alfv\'enic regimes respectively corresponding to run2 and run7 in Table \ref{table:simdata}. Other descriptions are the same as those in Figure \ref{fig:AM_no_FR}. 
} \label{fig:AMfreq}
\end{figure*}

In Figure \ref{fig:AMfreq}, we study how changes in radiation frequency affect AMs, according to run1 (sub-Alfv\'enic) and run6 (super-Alfv\'enic) in Table \ref{table:simdata}. With the purpose of covering observable frequency bands from the Low Frequency Array for radio astronomy (LOFAR) and the Square Kilometer Array (SKA), AM, as a function of the frequency from 0.01 to 100 GHz, is calculated. Through broadband radio frequency range of synchrotron polarization radiation, it is obvious that gradients of $I$, $Q$, $FR$, $P$, $P_{\rm v}$, $P_{\rm vg}$, $P_{\rm rad}$, $P_{\rm tang}$, and $\textit{\textbf{P}}$ are applicable to 
tracing the magnetic field of the ISM. More specifically, AMs of gradients of $P_{\rm rad}$ and $\textit{\textbf{P}}$ perform better in magnetic field tracing, compared with that of $I$, $Q$, $P$, $P_{\rm v}$, $P_{\rm vg}$ and $P_{\rm tang}$. It should be noticed that gradients of $FR$ and $I$ are frequency-independent; the former $FR$ is not an observable quantity. Especially in sub-Alfv\'enic simulation, gradients of $U$ and $QU$ can trace well magnetic fields at low frequencies, but another case at high frequencies. In super-Alfv\'enic simulation, however, they trace magnetic fields as well in low frequency regimes as in high frequency ones (see right panel). In addition, the curves of AM, especially for polarization vector method ($U/Q$), present trough-like shapes ranging from approximately 0.1 to 1 GHz. 
It may attribute to the strong Faraday depolarization with a large polarization deflection angle arising from the product of both Faraday rotation measure ($\phi$) and wavelength squared $\lambda^2$. Nonetheless, gradient techniques can trace magnetic fields better than traditional polarization method.

\subsection{Influence of mean magnetic field directions}\label{IMMFD}
In all data cubes listed in Table \ref{table:simdata}, there exist mean magnetic field directions along the horizontal ($x$-axis) direction. It is important to know whether synchrotron gradient techniques related to multifarious diagnostics can determine complex magnetic field configuration. Therefore, we intend to rotate real data cubes in Table \ref{table:simdata} by a certain angle to mimic more complex magnetic turbulence environment of the ISM. 

\begin{figure*}[t]
\centering
\includegraphics[width=0.45\textwidth]{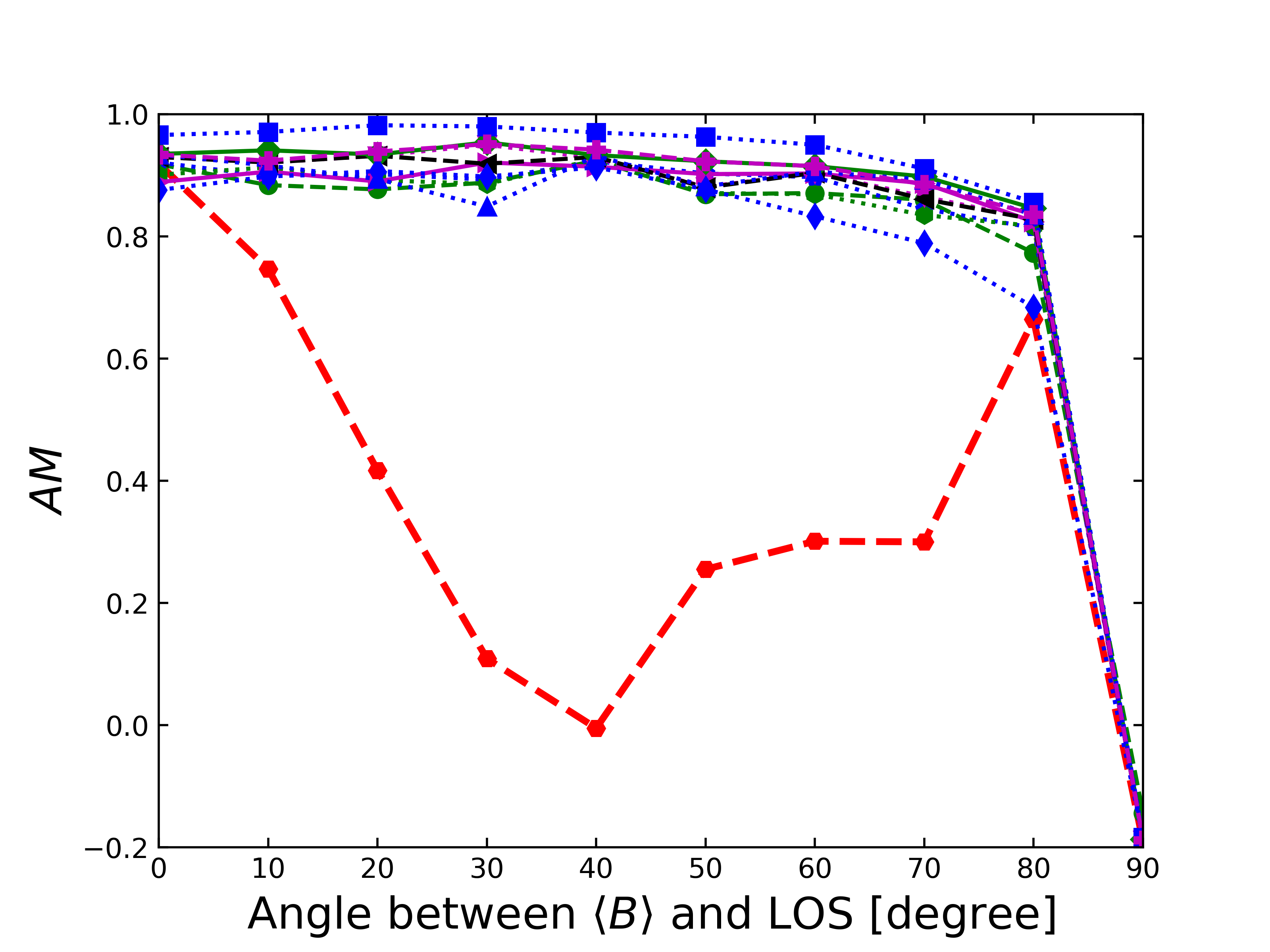}
\includegraphics[width=0.45\textwidth]{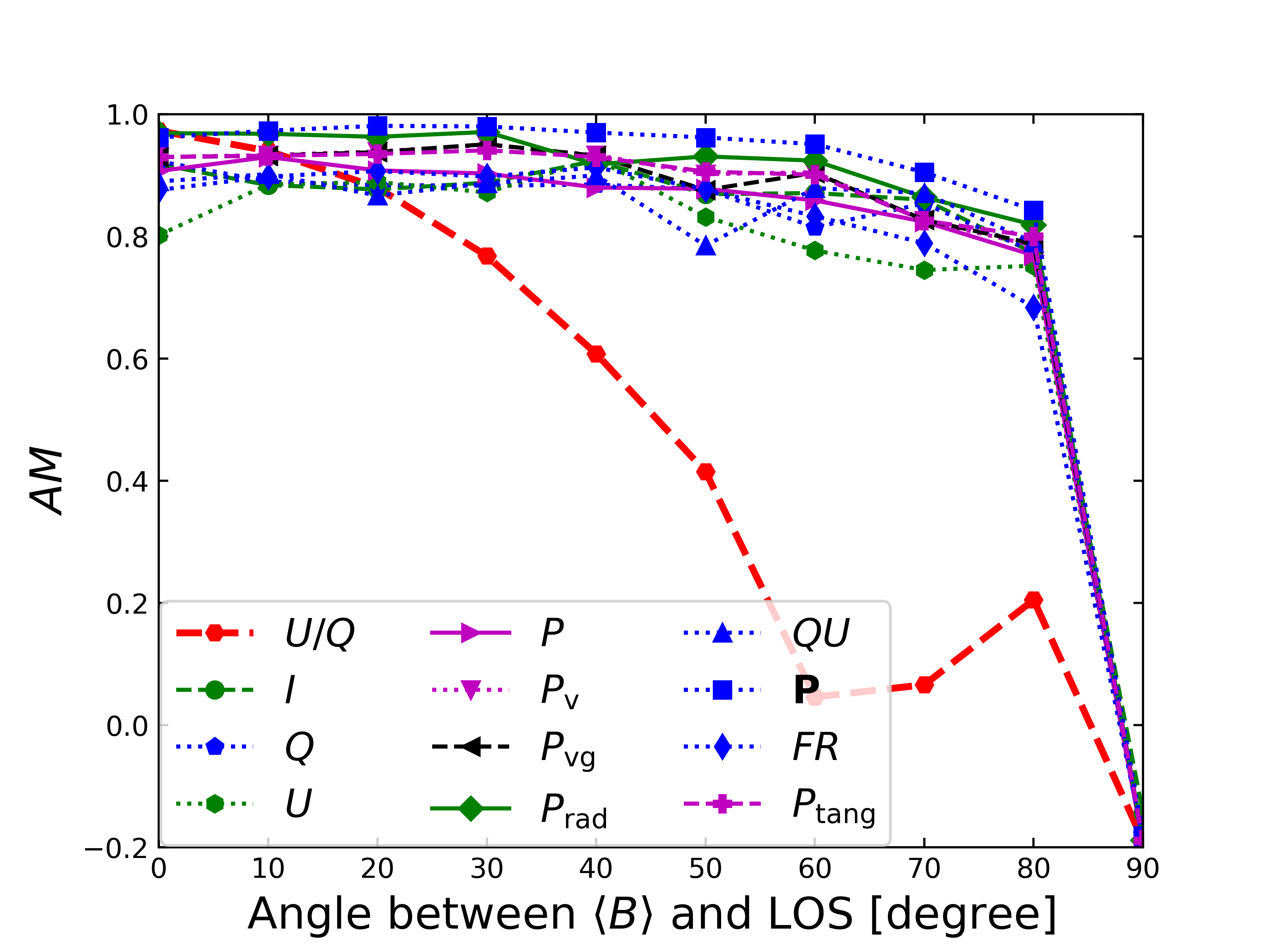}
\caption{AM, as a function of the angle of mean magnetic field directions to the LOS, between directions of mean magnetic fields and directions predicted by multifarious diagnostics, at frequencies $\nu=0.5$ (left panel) and 1 GHz (right panel), respectively.
} \label{fig:AMangle1}
\end{figure*}

\begin{figure*}[t]
\centering
\includegraphics[width=0.45\textwidth]{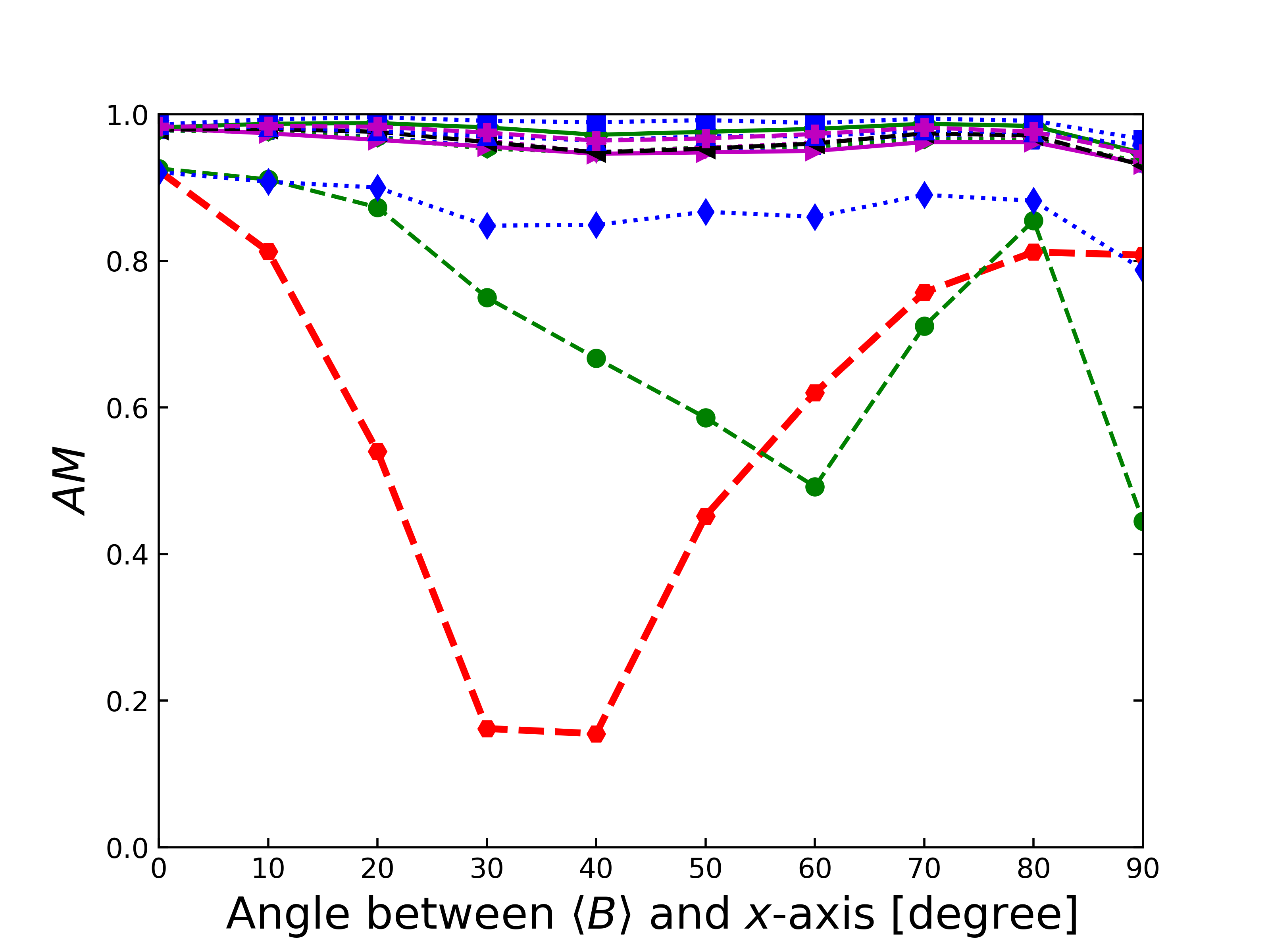}
\includegraphics[width=0.45\textwidth]{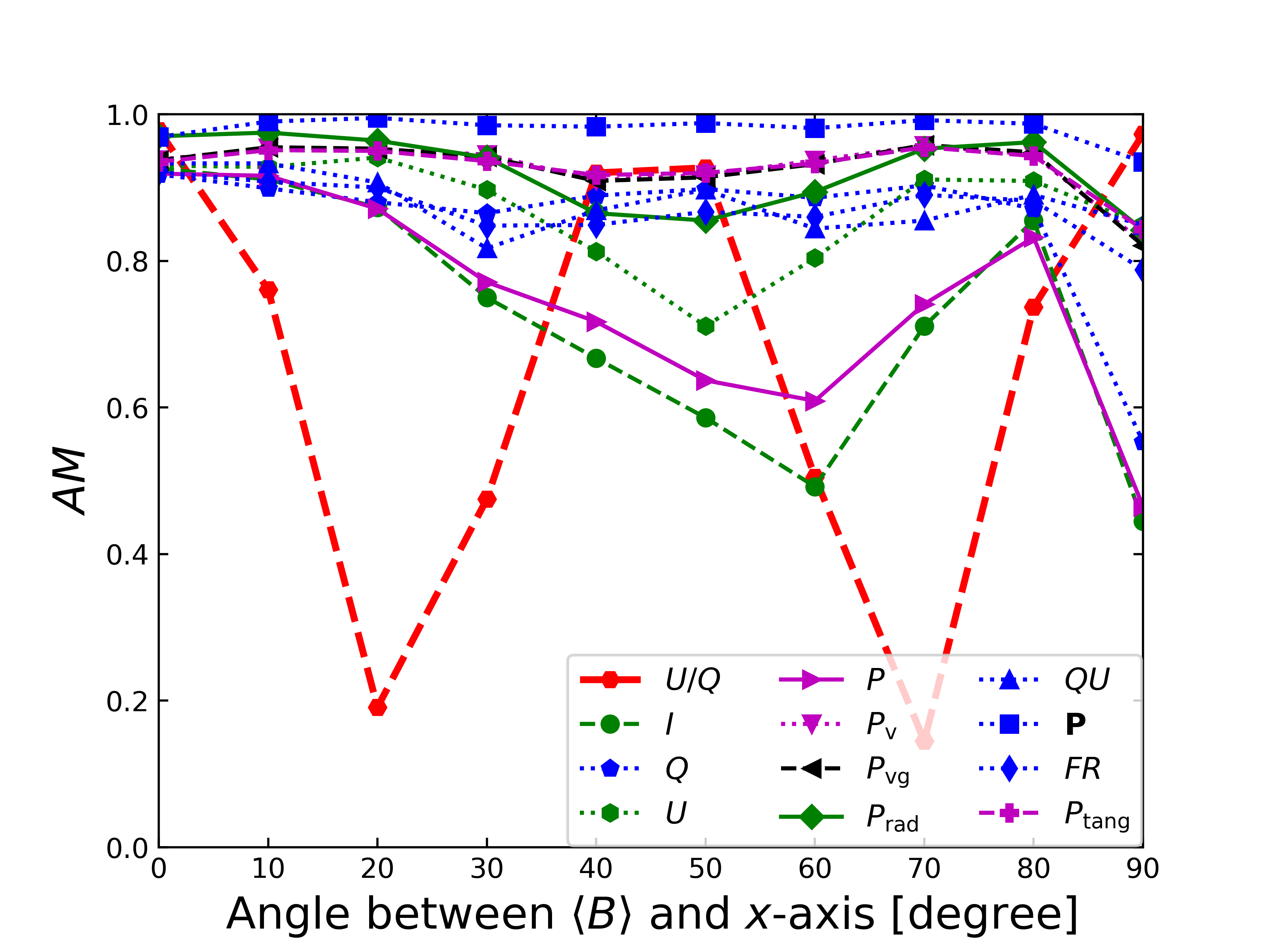}
\caption{AM, as a function of the angle of mean magnetic field directions to the horizontal direction ($x$-axis), between directions of mean magnetic fields and directions predicted by multifarious diagnostics, at frequencies $\nu=0.1$ (left panel) and 1 GHz (right panel), respectively.
} \label{fig:AMangle2}
\end{figure*}

\begin{figure}[t]
\centering
\includegraphics[width=0.5\textwidth]{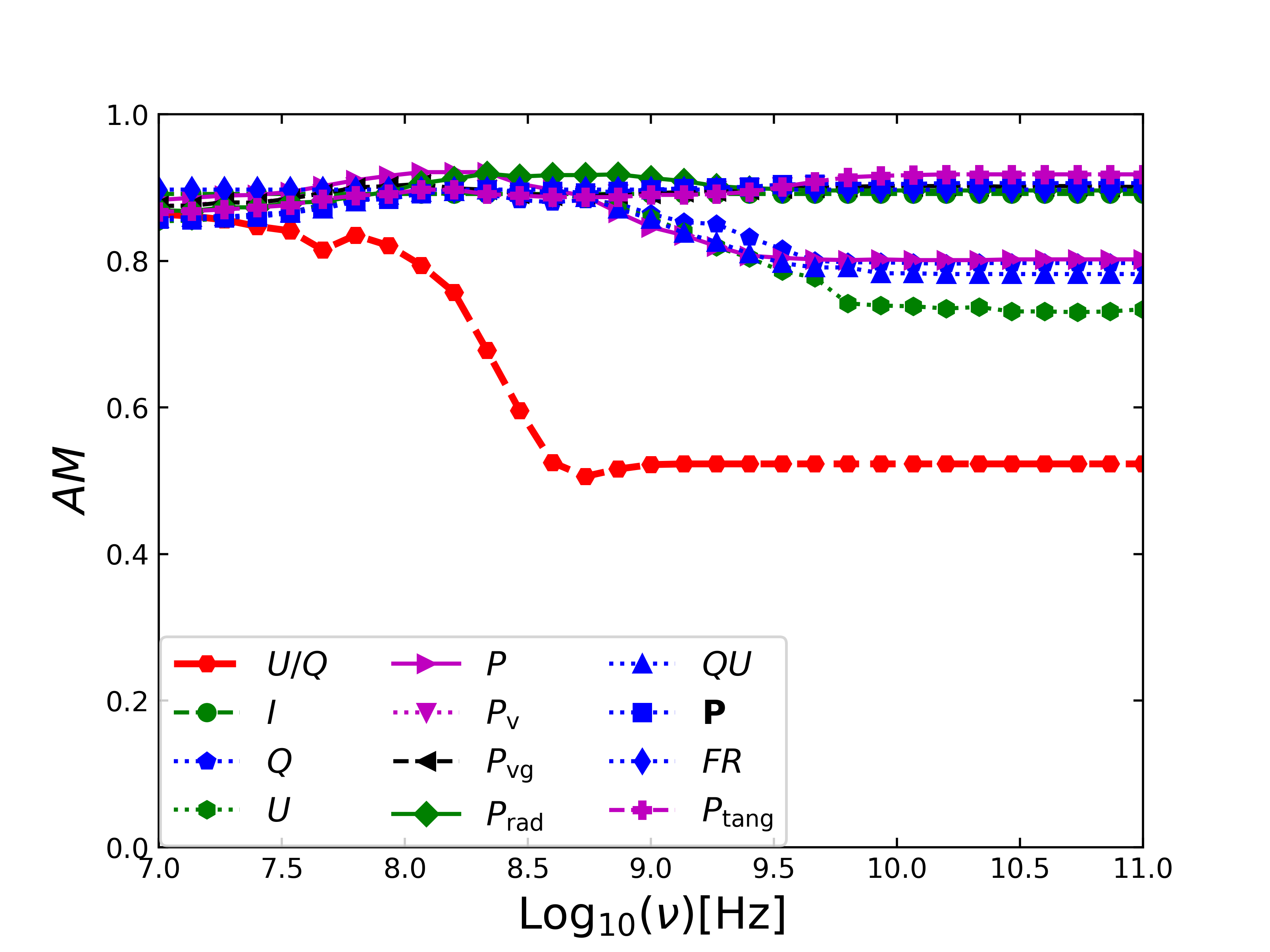}
\caption{AM, as a function of frequency, between directions of mean magnetic fields and directions predicted by multifarious diagnostics. With the back half of run2 in Table \ref{table:simdata} rotated by 30 degrees about the LOS and the front half left unchanged, data cubes are synthesized in this figure.
} \label{fig:AMangle3}
\end{figure}

\begin{figure*}[t]
\centering
\includegraphics[width=0.8\textwidth]{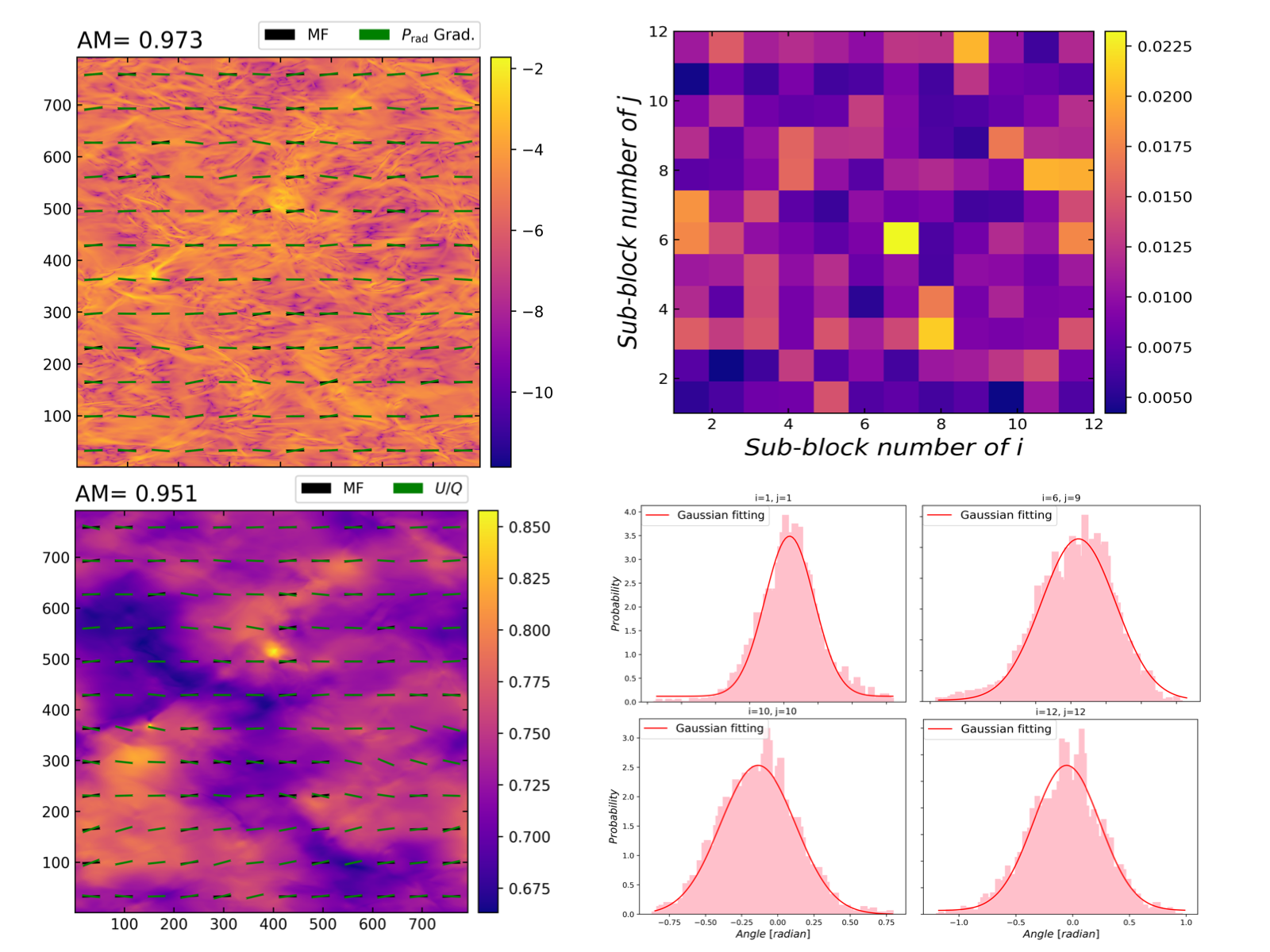}
\caption{ {\it Left upper panel}: Projected magnetic field (MF) directions vs. rotated $90^\circ$ gradient directions of $P_{\rm rad}$ with no Faraday rotation effect. In the background is the image of $P_{\rm rad}$ in units of mean synchrotron intensity on a logarithmic scale. {\it Right upper panel}: The peak error of Gaussian fitting for individual sub-block averaging corresponding to each sub-block region in the left upper panel. {\it Right lower panel}:  An illustration of Gaussian fitting of gradient statistics in four sub-block regions. {\it Left lower panel}: Projected magnetic field (MF) directions vs. rotated $90^\circ$ polarization vector directions by $\theta={1\over 2}{\rm arctan}(U/Q)$. In the background is the image of $P$ in units of mean synchrotron intensity. Simulations are based on run5 in Table \ref{table:simdata}.  } \label{fig:AM_res_noFR} 
\end{figure*}

In practice, we first rotate magnetic field and density data cubes labeled as run2 in Table \ref{table:simdata} by $\theta$ degrees about the $y$-axis, leading to the change of angle by the same $\theta$ between the mean magnetic field and the LOS. The rotated data enable us to obtain the relation between the AM and the rotated angle, as is plotted in Figure \ref{fig:AMangle1} at frequencies $\nu=0.5$ (left panel) and 1 GHz (right panel). The large AM values show the evidence of ideal alignment with exception for that of 90 degrees. The increase of the rotated angle brings into the decreased non-zero magnetic field components in the $x$-axis direction but gradually increased ones in the $z$-axis direction. When angles reach up to $\theta=90^\circ$, the $x$ component of mean magnetic fields approaches zero while the $z$ component becomes maximum. Consequently, no mean magnetic field exists in the plane of the sky to be traced by gradient techniques. Negative AM values (approximately $-0.2$) indicate that synchrotron gradient techniques should trace the projected random magnetic field distribution. It can be found that in this simulation the ability of traditional polarization vector method to determine magnetic field direction weakens.  

Then, data cubes marked as run2 are rotated by $\theta$ degrees about the LOS, resulting in the same $\theta$ degrees to the $x$-axis for mean magnetic field directions. Using these synthetized data, we calculate the AM between intrinsic mean magnetic field directions and magnetic field directions predicted by gradient techniques of multifarious diagnostics as a function of the rotated angle in Figure \ref{fig:AMangle2}, at frequencies $\nu=0.1$ and 1 GHz. As is shown in the figure, despite the reduced AMs emerging at different angles for $I$ at frequency 0.1 GHz and for $U$, $I$ and $P$ at frequency 1 GHz, most diagnostics of synchrotron gradients can provide large AM values for tracing mean magnetic fields of different orientations in the plane of the sky, with better AMs at $0.1$ GHz than at 1 GHz. The reduced AMs result from the fact that the diagnostic quantities are not rotationally invariant and the strong Faraday depolarization at low frequency breaks dependence on the coordinate system. Naturally, the translationally and rotationally invariant quantities of synchrotron emission are important for determining directions of mean magnetic fields.

Furthermore, the way that only the back half part of run2 is rotated by 30 degrees about the LOS and the front half remains the same helps us synthetize new data cubes. We plot in Figure \ref{fig:AMangle3} the AM distributions of multifarious diagnostics when changing the radiation frequency. In the low ($\lesssim 1$ GHz) frequency  range, multifarious diagnostics of synchrotron gradients provide large AM values to reliably trace directions of complex mean magnetic fields. In the high ($\gtrsim1$ GHz ) frequency range, some of diagnostics, such as non-translational and non-rotational $P$, $Q$, $U$ and $QU$, present slightly reduced AMs of $\sim 0.8$.  Accordingly, it is necessary for robustly tracing directions of complex mean magnetic fields by the synergies of multifarious diagnostics. Apparently, synchrotron gradient techniques can trace the directions of mean magnetic fields with complex configuration.

\begin{figure}[t]
\centering
\includegraphics[width=0.45\textwidth]{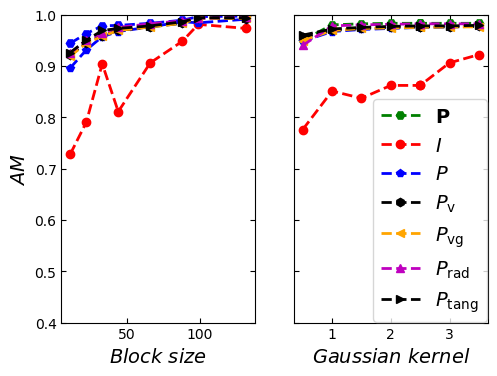}
\caption{Influence of sub-block size (left panel) and Gaussian kernel (right panel) on the AM at the frequency of $\nu=0.1$ GHz, according to run1 in Table \ref{table:simdata}. The adopted parameters are a Gaussian kernel of $\sigma=2$ (left) and sub-block size of 66 pixels (right). 
} \label{fig:block_kernel} 
\end{figure}

\subsection{Comparison of multifarious diagnostic techniques} \label{CVDT}
In Section \ref{MultiFM}, we explored the flexibilities of gradient measures of multifarious synchrotron diagnostics to determine magnetic field directions. Just with naked eyes can we find that gradients of $\textit{\textbf{P}}$ and $P_{\rm rad}$ have an advantage over those of $I$, $Q$, $P$, $P_{\rm v}$, $P_{\rm vg}$ and $P_{\rm tang}$ in magnetic field tracing. On the basis of all data cubes listed in Table \ref{table:simdata}, the precise comparison is made in the AM performance of individual multifarious diagnostics on average of 31 AM values in the frequency ranging from 0.01 to 100 GHz. Consequently, as the Alfv\'enic Mach number increases, the averaged AM values slightly decrease. However, the AM values greater than 0.9 of such diagnostic quantities as $I$, $P$, $P_{\rm v}$, $P_{\rm vg}$, $P_{\rm rad}$, $P_{\rm tang}$ and $\textit{\textbf{P}}$, still can be achieved even $M_{\rm A}=1.71$. It is evident that all the above effective diagnostics, among which, $\textit{\textbf{P}}$ and $P_{\rm rad}$ provide larger AMs for each simulation, can be employed to robustly trace magnetic fields.

\begin{table*}[t]
\caption {Averaged AM between mean magnetic field orientations and directions predicted by multifarious diagnostics. The AM values in the AMrun1 row are obtained using data cubes marked as run1 in Table \ref{table:simdata}, and so on so forth. Each averaged AM value listed in this Table is determined by averaging several AM values that are calculated at frequencies ranging from 0.01 to 100 GHz in 30 logarithmic intervals. }
 \centering
 \begin{tabular}{c c c ccccccccccc}%
 \hline
    AM      &$U/Q$  & $ I $  &  $ Q$   &    $U $  &  $ P$   &  $P_{\rm v}$  &  $P_{\rm vg}$ & $ P_{\rm rad}$  & $QU$  & $\textit{\textbf{P}}$ & $FR$  &  $P_{\rm tang}$ \\ \hline \hline
AMrun1 & 0.978  & 0.895 & 0.937 &  0.739 & 0.941 &0.961 &0.962 &0.980 &0.758 &0.983 &0.912 & 0.949 \\ 
AMrun2 & 0.953  &0.918 & 0.926 &  0.690 & 0.918 &0.963 &0.961 &0.972& 0.706 &0.976 &0.876 &0.947 \\
AMrun3 & 0.932  &0.932 & 0.929  & 0.781 & 0.919& 0.961 &0.958& 0.973& 0.797 &0.976 &0.876 &0.938\\
AMrun4  &0.931 & 0.946 & 0.951 &  0.844 & 0.947 &0.965 &0.964 &0.978& 0.823& 0.980& 0.903& 0.948\\
AMrun5 & 0.895 & 0.908 & 0.928  & 0.856 & 0.928 &0.955 &0.954 &0.968 &0.865& 0.968 &0.916& 0.936\\
AMrun6 & 0.871  &0.895 & 0.908 &  0.854&  0.922& 0.948&0.948& 0.963 &0.892 &0.964 &0.895 &0.908\\
AMrun7 & 0.860 & 0.929 & 0.931  & 0.880  &0.939 &0.954 &0.953 &0.964& 0.913& 0.964 &0.940 &0.944\\
AMrun8 & 0.792  &0.929  &0.911 &  0.882 & 0.930 &0.940& 0.939& 0.951 &0.901 &0.944& 0.879 &0.921\\
AMrun9  &0.749  &0.866 & 0.896  & 0.852 & 0.902& 0.928& 0.926& 0.936 &0.881 &0.930 &0.915& 0.913\\
AMrun10 &0.680 & 0.922&  0.890 &  0.863 & 0.902 &0.914 &0.913 &0.920 &0.881 &0.917 &0.898 &0.900\\ \hline \hline
\end{tabular}
 \label{table:AMsim}
\end{table*}

\subsection{Reliability of tracing magnetic field}
As is described in Section \ref{GMT}, the Gaussian fitting peak of a sub-block, i.e., sub-block averaging is used to determine the maximum gradients of diagnostic quantities. To exemplify more details of each sub-block, we plot the directions of projected magnetic fields (in black) and the gradient directions of $P_{\rm rad}$ (in green), only to find a large AM=0.973, shown in the left upper panel of Figure \ref{fig:AM_res_noFR}. The sub-block size of $66^2$ pixels accounting for the entire $792^2$ pixels, the whole map is divided into $12^2$ sub-blocks with the image of $P_{\rm rad}$ in the background in units of mean synchrotron intensity on a logarithmic scale. The right upper panel of Figure \ref{fig:AM_res_noFR} reveals the reliability of the Gaussian fitting through the peak angle error in units of radian with $95\%$ confidence level for each sub-block. What can be clearly seen is that the smaller colorbar value goes with the higher fitting confidence. Meanwhile, the right lower panel of this figure, providing the Gaussian fitting of 4 sub-blocks for gradients of $P_{\rm rad}$, illustrates good distributions of gradients. A good comparison can be made between the results presented in the left upper and left lower panel of Figure \ref{fig:AM_res_noFR} for new gradient techniques and traditional polarization method accordingly in tracing magnetic fields.

In the above calculation, AM values are obtained by means of a Gaussian kernel of $\sigma=2$ and the sub-block size of 66 pixels. In terms of the AM as a function of the sub-block size (left panel) and Gaussian kernel (right panel), other explorations are also made in Figure \ref{fig:block_kernel} for certain effective diagnostics (see Section \ref{CVDT}) at the frequency of 0.1 GHz. Clearly, a continual growth of AM value can be noticed with the increase of sub-block size except for the almost stability with the block size up to 66 pixels. Similar trend can be found with AM as a function of Gaussian kernel: AM value grows as Gaussian kernel increases but remains steady when the kernel value moves greater than 1. Therefore, what can be inferred is that parameters $\sigma=2$ and 66 pixel sub-block size we used in this paper are ideal for the current numerical resolution and simulated radiation frequency coverage.

\begin{figure*}[t]
\centering
\includegraphics[width=0.8\textwidth, bb=55 55 650 530]{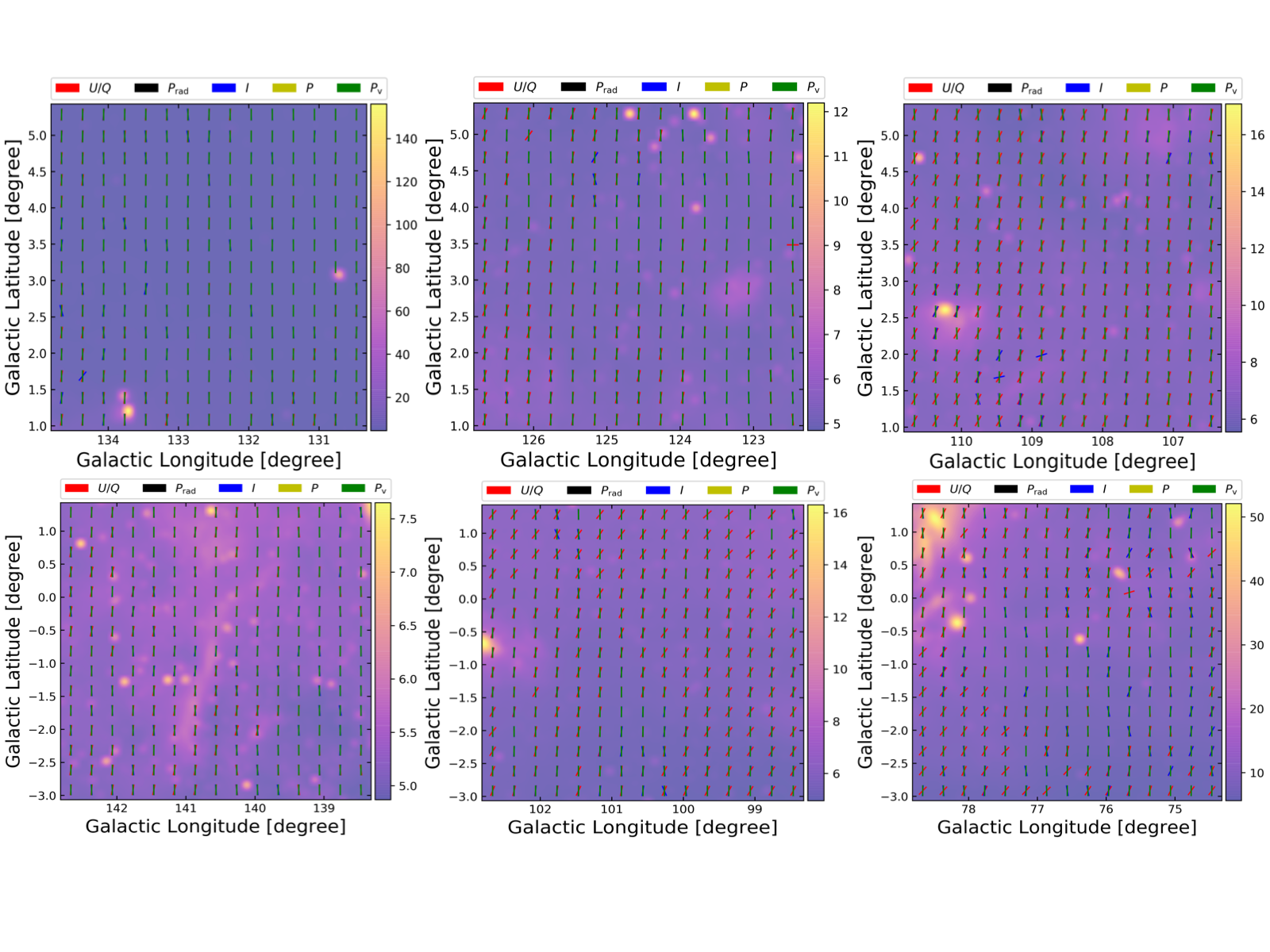}
\caption{Magnetic field directions within several patches of the Galactic disk predicted by gradient techniques with the CGPS data at $1.42\rm\ GHz$. Magnetic field orientations (in red segment line) traced by synchrotron polarization vector is shown in each panel for comparison. The images in the background display the synchrotron intensities in units of brightness temperature. } \label{fig:AM_map_CGPS} 
\end{figure*}

\section{Application to observation}\label{AtoPo}

\begin{table*}[t] %
\caption {AM between directions predicted by gradients of multifarious diagnostics and directions by traditional polarization vector. The first row AMMX1 
are obtained using the observational data from the MX1 patch of the sky plane presented in Figure 4 of \cite{Taylor03}, and so on.}
 \centering
 \begin{tabular}{ccccccccccc}%
 \hline
  AM [Longitude (deg), Latitude (deg)]     &$P_{\rm rad}$ &   $I$ &  $P$  &    $P_{\rm v}$ &  $P_{\rm vg}$ &  $P_{\rm tang}$ &  $\textit{\textbf{P}}$ & $QU$ &  $Q$  &  $U$\\ \hline \hline
AMMX1 [$138.8<l<134.3$, $-3.1<b<1.4$]   &   0.961 & 0.918  &0.961  &0.961 &0.961 & 0.961  & 0.961  & 0.961 & 0.961 &0.961\\
AMMO1 [$78.8<l<74.3$, $-3.1<b<1.4$]   &  0.717 & 0.693 & 0.717  &0.717 &0.717 &0.717 & 0.717  & 0.717 & 0.717 &0.717\\
AMMH1 [$102.8<l<98.3$, $-3.1<b<1.4$]    & 0.625 & 0.638 & 0.625 & 0.625 & 0.625 & 0.625 & 0.625 &  0.625 & 0.625 & 0.625\\
AMMW1 [$142.8<l<138.3$, $-3.1<b<1.4$]    & 0.966 & 0.951 & 0.966 & 0.966 &0.966 & 0.966 & 0.966  & 0.966 & 0.966 &0.966 \\
AMMM1  [$86.8<l<82.3$, $-3.1<b<1.4$]    &0.436 & 0.438 & 0.436 & 0.436 &0.436 &0.436 & 0.436  & 0.436 & 0.436 &0.436 \\
AMMF2 [$110.8<l<106.3$, $0.9<b<5.4$]      &0.679  &0.702 & 0.682 & 0.682 & 0.684 & 0.682 & 0.676 &  0.685 & 0.693 &0.688\\
AMMB2 [$126.8<l<122.3$, $0.9<b<5.4$]      &0.927 & 0.927 & 0.927 & 0.927 & 0.927& 0.927&  0.927 &  0.927  &0.927& 0.927\\
AMMY2 [$134.8<l<130.2$, $0.9<b<5.4$]    & 0.988 & 0.973 & 0.988 &  0.988 & 0.988 & 0.988 & 0.988 &  0.988 & 0.988 &0.988 \\ \hline \hline         
\end{tabular}
 \label{table:AMobs}
\end{table*}

Until now, gradient tracing techniques for multifarious diagnostics of synchrotron emission are explored by using simulation data. AM obtained in Section \ref{nums} is good enough to motivate us to study magnetic field properties of the ISM within the Milky Way, by using the archive data from the CGPS at $1.42\rm\ GHz$. The CGPS is a project integrating radio, millimeter, and infrared surveys of the Galactic plane to obtain arc minute-scale images of all major components of the ISM over a large part of the Galactic disk. Generally, the synchrotron radio surveys conducted at the Dominion Radio Astrophysical Observatory (DRAO), i.e., the polarization observations for the low-frequency component of the CGPS provided us the merged high-resolution database (\citealt{Taylor03}). The DRAO Synthesis Telescope surveys imaged a 73-degree part of the Galactic plane (with longitude extent 74.2 $< l <147.3$ degrees and latitude extent $-3.6 <b< +5.6$ degrees) between 1995 April and 2000 June. The DRAO fields were combined into $1024\times1024$ pixel mosaic images with near $1'$ resolution on a Galactic Cartesian grid, covered by 36 mosaics ($5^\circ.12\times 5^\circ.12$ for each) in Figure 4 of \cite{Taylor03}. In the current paper, we randomly select eight mosaics of them to study magnetic field directions of the ISM and only extract $900\times900$ pixels from $1024\times1024$ pixel mosaic images to avoid the margin of the images.

Observational data are different from numerical simulation one where the potential direction of magnetic fields is set in advance. For latter one, magnetic field orientations predicted by gradient techniques can be compared with intrinsic magnetic field directions in data cubes. Using the CGPS archive data selected and Gaussian kernel of $\sigma=2$ pixels to filter noise-like structure of the data, we provide in Table \ref{table:AMobs} AM values concerning alignment of rotated $90^\circ$ gradient directions of multifarious synchrotron diagnostics against traditional polarization method.\footnote{In this paper, the AM from simulation data is determined by alignment of directions of underlying magnetic field to directions traced by gradients of synchrotron diagnostics, while the AM from observational data is decided by alignment of directions by traditional polarization method to directions by gradients of synchrotron diagnostics.} From this table, gradients of multifarious diagnostics for each patch of the Galactic disk are found able to predict magnetic field directions consistently. What is interesting is the tracing capability that $QU$ and $U$ display in this table but fail to present in numerical testing in high frequency regime. By and large, high consistency can be seen in the AM of several diagnostic gradient measurements except for marginal differences for that of synchrotron intensity $I$. It should be attributed to the fact that the polarized emission suffers from an influence of a foreground Faraday screen, which is comprised of the magnetization turbulent ISM acting on the emission from the Galactic background synchrotron.

Besides, to understand the distributions of magnetic field directions in each sub-block, we show in Figure \ref{fig:AM_map_CGPS} the predicted results for six patches of the Galactic disk with the diagnostic quantities $I$, $P$, $P_{\rm rad}$ and $P_{\rm v}$ as an example. As is shown in the left panels, the rotated $90^\circ$ gradient directions align well to the rotated $90^\circ$ electric vector directions by polarization vector method through the entire image region. In other panels, a good alignment between them occurs only in some sub-block regions. Images in the background demonstrate synchrotron intensities in units of brightness temperature. Satisfying alignment is supposed to happen in the weak or no Faraday depolarization region where the rotated $90^\circ$ electric vectors by polarization vector can determine the direction of magnetic fields, while the alignment appears to be unsatisfying in the significant Faraday depolarization region where polarization vector method cannot work. For the latter, the direction of magnetic field tracing by traditional polarization method is not a real direction due to depolarization effect. According to the AM value, we can estimate the extent of Faraday rotation depolarization, that is, the smaller the AM is, the larger the deflection angle of electric vectors is. More importantly, gradient directions of multifarious diagnostics plotted in segment lines provide prediction for turbulent magnetic field directions of the ISM within the Galactic disk.
  
Obviously, the complexity of the Faraday depolarization will not impede us from measuring turbulent interstellar magnetic fields of the Milky way. We expect that when small-scale structures can be well resolved with the advent of a higher resolution observation, the potential small-scale turbulence should again lead to the synchrotron gradient alignment as to the projected magnetic fields. If so, the synergies of gradients of multifarious diagnostics can be employed to more accurately trace magnetic field directions in the ISM of the Milky way, and the synergies of different methods can further enhance the reliability of measurement.

\section{Discussion}
Gradients of both synchrotron intensity (SIGs) and polarization intensity (SPGs, LY18; \citealt{Zhang19}) have been adopted recently in the study of magnetic field direction determination. Other possibilities in tracing magnetic fields are explored in this paper, such as synergies of synchrotron emission gradient techniques. High resolution numerical simulations indicate that gradients of some extra diagnostics, such as $Q$, $\textit{\textbf{P}}$, $P_{\rm v}$, $P_{\rm vg}$, $P_{\rm rad}$ and $P_{\rm tang}$, are also applicable in magnetic field direction tracing. To verify the reliability of the techniques, we have further applied new techniques in combination with real observational data from the CGPS to the studies of interstellar magnetic fields of the Milky Way. In practice, gradient statistics of the diagnostics mentioned above are highlighted to constrain projected magnetic field directions. However, other properties of interstellar turbulence, e.g., Alfv{\'e}nic Mach number, can also be determined by these diagnostics and will be discussed somewhere else.

As is shown in Figure \ref{fig:Polnoise}, it is apparent that many small-scale noise-like structures resulting from the strong Faraday depolarization can be seen in the map of polarization intensities at low frequencies. In addition, numerical noise is inevitable in the process of the first-order derivative, second-order derivative, or polynomial operation to the data, especially in other more complex diagnostics such as the gradients for $P_{\rm v}$, $P_{\rm vg}$, $P_{\rm rad}$ and $P_{\rm tang}$. According to our numerical experience, the removal of the numerical noise of post-gradient data by means of the Gaussian filter is necessary to improve the alignment determination. Therefore, synergies of multifarious statistical diagnostics in combination with noise removal techniques can optimize our results with no doubt. Practically, we smoothed the noise of simulation and observational data by using the Gaussian filter to enhance the measurement accuracy of synchrotron gradients. It will be of great necessity to further examine the influence of noise on different diagnostics. Take gradients of maximum amplitudes of the radial component of the directional derivative (see Equations (\ref{eq:prad})) for example, the results can be noise-influenced for the radial component is translationally invariant.

The current work is involved in the AM for gradients of several statistical diagnostics with the consideration of the spatially coincident synchrotron emission and Faraday rotation regions. We know that the Stokes parameters $Q$ and $U$ are orthogonal decompositions of a linear polarization information. Here, $Q$ describes vertical and horizontal polarization, while $U$ measures the polarization along diagonals at $45^\circ$ and $135^\circ$ to the horizontal direction (\citealt{Stokes1851}). Despite its tracing in magnetic field directions, the gradient of $Q$ may result in an incomplete statistic due to a lack of diagonal direction statistics. Furthermore, some of gradient diagnostics such as $Q$, $U$, $I$, $P$ and $QU$, which are not translationally and rotationally invariant, cannot work well in certain simulations. Therefore, we point out that the rotationally invariant quantities are crucial to tracing mean magnetic field directions.

As is mentioned in Section \ref{MuoMT}, the statistics of gradient techniques reflect the statistics of the underlying anisotropic turbulence. The gradient techniques would provide a better way tracing magnetic fields provided that fluctuations at small scales dominate over large scales. As is seen in Figures \ref{fig:AM_no_FR} and \ref{fig:AM_with_FR}, the AM decreases with the increasing Alfv{\'e}nic Mach number. As for the super-Alfv{\'e}nic turbulence, the motion of eddies on a large scale is rarely limited by magnetic fields, so the hydrodynamic processes control turbulence interactions. The AM would be increased when the spatial structure at large scales is removed (see \citealt{Zhang19} for the SPGs). It should be interesting to find other techniques for studying the structure of large-scale magnetic fields.

In observational studies of the Galactic turbulent magnetic fields, a good alignment measure (between rotated $90^\circ$ gradient directions and polarization vectors) might well be found in most of the regions at the high latitude, although less alignment may occur as the latitude decreases. On the one hand, the Galactic synchrotron halo emits synchrotron polarized radiation without much of Faraday rotation for the low density of thermal electrons, whereas the Galactic disk plane may experience substantial Faraday rotation that results in alignment reduction. On the other hand, the decrease of AM is due to the existence of the poorly resolved synchrotron structures not directly associated with a turbulent fluctuation, under which circumstances our gradient techniques seem not that sufficient. What's more, the potential complication may emerge in the Galactic disk plane when it comes to the distant regular structure (e.g., supernova remnant), leaving the gradient measurements distorted for their reliance on the resolved turbulence associated with the nearby objects. Therefore, our results presented in this paper on the basis of the CGPS data should face up to future testing from higher resolution LOFAR and SKA data cubes. More effort should be made in the development of statistical techniques for the analysis of polarization diagnostic gradients, which will be beneficial to accurately identify the properties of magnetic fields in interstellar turbulence such as magnetic field strength and compressibility.

\section{Summary}
In this paper, we have used the synergies of gradients of synchrotron emission diagnostics to trace magnetic field directions, based on the data from high resolution numerical simulation and the CGPS archive. The resultant findings are listed as follows.

\begin{enumerate}[wide, labelwidth=!, labelindent=1pt]
\item The rotationally invariant quantities of synchrotron gradient techniques are highly required for tracing directions of mean magnetic fields. However, the non-rotationally invariant quantities are found to be independent from the coordinate system in the case of the low-frequency strong Faraday rotation effect.

\item Magnetic field directions are found able to be successfully traced by gradients of several diagnostics of synchrotron emission, including $I$, $P$, $Q$, $\textit{\textbf{P}}$, $P_{\rm v}$, $P_{\rm vg}$, $P_{\rm rad}$ and $P_{\rm tang}$. Among which, $P_{\rm rad}$ should be the best choice for tracing mean magnetic fields of the Galactic ISM. 

\item Higher reliability of measurement can be definitely achieved through the employment of the synergy of synchrotron diagnostics in mean magnetic field directions, where low-frequency and strong Faraday depolarization does not impede us from directions tracing for complex magnetic field configuration.

\item Through application of synchrotron gradient techniques with the CGPS data, the synergies of synchrotron gradient techniques are shown to be more feasible than traditional polarization measurement in the low-frequency strong Faraday rotation regime, providing better predictions for magnetic field directions of the ISM within the Galactic plane.

\item Being low-frequency-regime-oriented, the development of synchrotron emission gradient techniques paves a way for the application of these new techniques to the LOFAR and SKA data cubes. 
\end{enumerate}

\acknowledgments
We thank the anonymous referee for constructive comments and suggestions that significantly improved this paper. J.F.Z. thanks the supports from the National Natural Science Foundation of China (grant Nos. 11973035 and 11703020) and the Hunan Provincial Natural Science Foundation (grant No. 2018JJ3484). A.L. acknowledges the support of NSF grant AST 1816234 and 1715754.


\end{document}